\documentclass[12pt,ascmac]{report}
\usepackage{eclepsf}
\usepackage{tascmac}
\usepackage{url}
\usepackage{amsmath}
\usepackage[numbers]{natbib}
\usepackage{graphics}
\usepackage{graphicx}
\usepackage{color}
\usepackage{indentfirst}
\usepackage{lscape}
\usepackage{fancyhdr}

\usepackage{hyperref}
\usepackage{CJKutf8}

\begin{document}

\pagenumbering{roman}

\begin{titlepage}
    \begin{center}
        \begin{large}
            Master Thesis (Academic Year 2018)\\
            \vspace{2em}
            ~ \bigskip ~ \\
            \LARGE {\bf Blockchain Storage Load Balancing Among DHT Clustered Nodes\\}
            ~ \bigskip ~ \\
            \vspace{1em}
        \end{large}

        \vspace{5em}
        \begin{figure}[h]
            \centering
            \includegraphics[width=6cm]{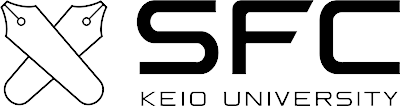}
        \end{figure}
        \vspace{5em}

        \begin{center}
            \noindent \large Keio University\\ Graduate School of Media and Governance
        \end{center}

        \begin{center}
            \noindent \LARGE Ryosuke Abe\\
        \end{center}
        \begin{center}
            \noindent \large Main Supervisor Jun Murai\\
            \noindent \large Co-Supervisor Osamu Nakamura\\
            \noindent \large Co-Supervisor Shigeya Suzuki\\
        \end{center}

    \end{center}
\end{titlepage}

\pagenumbering{Roman}

\begin{center}
    \noindent Abstract of Master's Thesis - Academic Year 2018
\end{center}

~ \bigskip ~ \\

\begin{center}
    \begin{large}
       Blockchain Storage Load Balancing\\Among DHT Clustered Nodes
    \end{large}
\end{center}

~ \bigskip ~ \\
    In Bitcoin, to independently verify whether new transactions are correct or not, a type of a node called ``Full Node'' has to hold the whole of historical transactions. The transactions are stored in ledger called ``Blockchain.''  Blockchain is an append-only data structure. Thus, to operate Full Nodes, the required storage capacity would grow too large for resource-constrained devices. Due to the limitation, the existing lightweight node scheme is that a node relies on other Full Nodes. In this thesis, to reduce storage capacity with keeping the independence of each node, we propose a storage load balancing scheme ``KARAKASA'' using Distributed Hash Table (DHT). In KARAKASA, nodes distributedly keep the whole blockchain among DHT networked nodes.  We evaluated KARAKASA from the view of storage capacity and independence. As a result, a node in a cluster does not need to trust other nodes. We concluded that nodes in a DHT cluster can behave like Full Nodes without holding the whole blockchain.
~ \bigskip ~ \\

Keywords : \\
\underline{1. Bitcoin},
\underline{2. Storage},
\underline{3. Distributed Hash Table},
\underline{4. Blockchain}

~ \bigskip ~ \\
\begin{flushright}
    Graduate School of Media and Governance, Keio University\\
    ~ \\
    \begin{large}
        Ryosuke Abe
    \end{large}
\end{flushright}

\clearpage

\begin{CJK}{UTF8}{min}
\begin{center}
    \noindent 修士論文要旨 2018年度 (平成30年度)
\end{center}

~ \bigskip ~ \\
\begin{center}
    \begin{large}
        分散ハッシュテーブルを用いた\\Blockchainストレージのロードバランシング
    \end{large}
\end{center}

~ \bigskip ~ \\
    Bitcoinでは，新規取引の正当性を独立して検証するために， 「Full Node」と呼ばれるタイプのノードは過去の取引データを全て保持する．取引データは，「Blockchain」と呼ばれる台帳に保存される．Blockchainは，追記のみ可能なデータ構造である．そのため，ノードをFull Nodeとして動作させるために必要とされるストレージ容量は肥大化し続ける．したがって，ストレージ容量に制約のあるデバイスをFull Nodeとして動作させることは困難である．また，既存の軽量ノードは，検証のために他のFull Nodeに依存しなければならない．本論文では，各ノードの独立性を保ちながら必要とされるストレージ容量を軽減するために，分散ハッシュテーブル（DHT）を用いた負荷分散方式「KARAKASA」を提案した． KARAKASAでは，ノードはDHTクラスタ内でBlockchain全体を分散的に保存する．提案手法の評価として，ストレージ容量と独立性の観点から分析を行なった．その結果，クラスタ内のノードは他のノードを信頼する必要はなく，DHTクラスタ内のノードは，Blockchain全体を保存せずともFull Nodeと同様に動作可能である．
~ \bigskip ~ \\

キーワード \\
\underline{1.Bitcoin}，
\underline{2.ストレージ}，
\underline{3.分散ハッシュテーブル}，
\underline{4.Blockchain}，

\begin{flushright}
    慶應義塾大学大学院 政策・メディア研究科\\
    ~ \\
    \begin{large}
        阿部 涼介
    \end{large}
\end{flushright}
\end{CJK}
\clearpage

\tableofcontents
\clearpage

\listoffigures
\clearpage

\listoftables
\clearpage

\pagenumbering{arabic}
\pagestyle{fancy}
\fancyhead{}
\fancyhead[RO,LE]{\leftmark}

\chapter{Introduction} 
Bitcoin~\cite{bitcoin} is a peer-to-peer electronic payment system. Bitcoin was invented by ``Satoshi Nakamoto\footnote{Anonymous individuals or groups who call them ``Satoshi Nakamoto.''}'' in 2009. In Bitcoin, each node store transactions in a public ledger. The public ledger is called ``Blockchain.'' Each node can verify the correctness of new transactions by referring to the blockchain stored in local storage. Each node stores verified transactions in the blockchain in the local storage. Each node stores transactions as a unit ``Block.'' Block includes verified transactions.

Typically, to verify new transactions, a type of Bitcoin node called ``Full Node'' has to hold the whole blockchain. Full Nodes can verify all new transactions locally. Thus, each Full Nodes can behave without depending on other nodes.

\begin{figure}[t]
  \centering
  \includegraphics[width=14cm, bb=0 0 461 346]{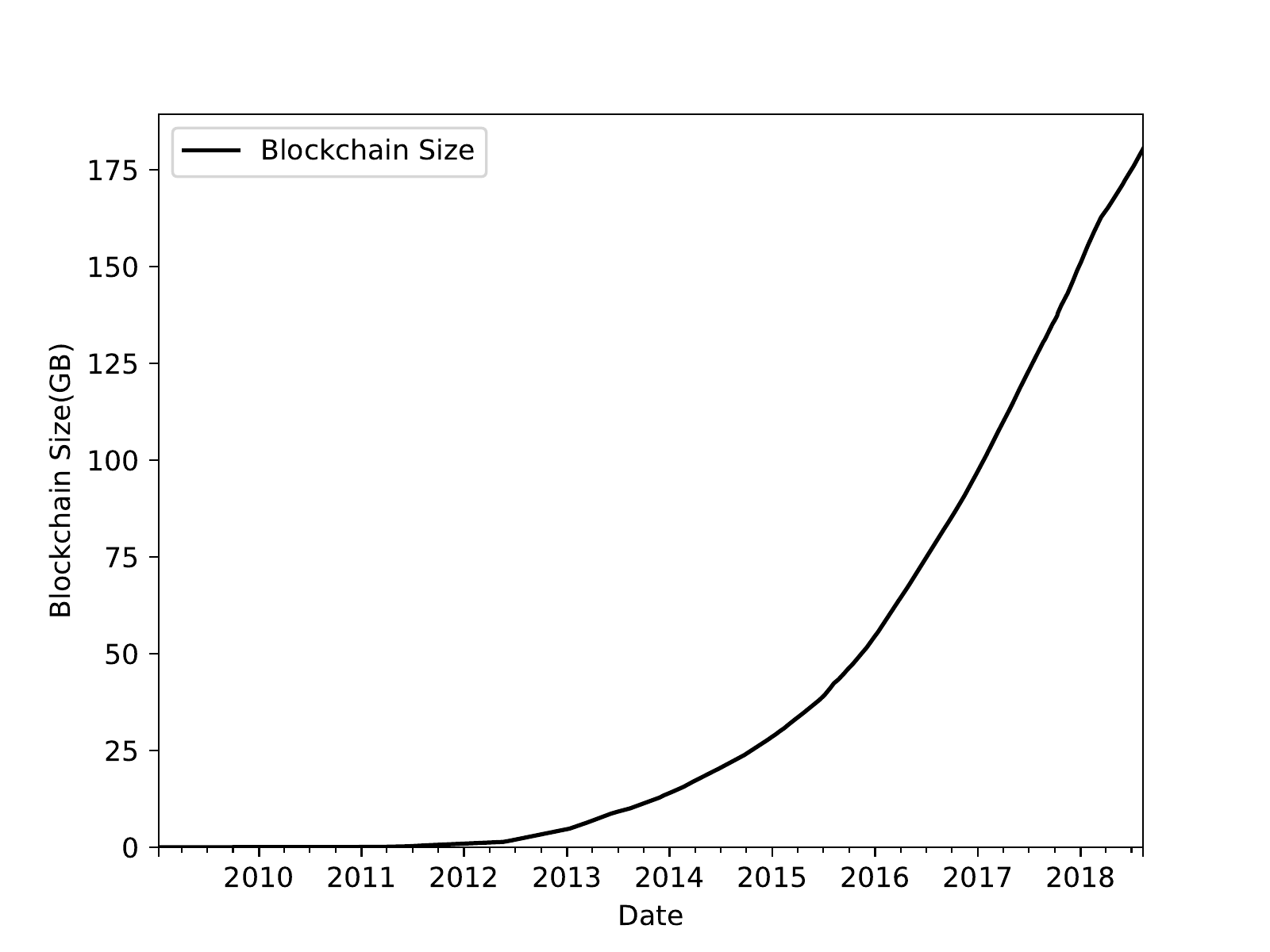}
  \caption{Bitcoin Blockchain Size Grows}
  \label{img:btcchaingrows}
\end{figure}

To behave as a Full Node, each node requires enough amount of storage. The blockchain is append-only data storage scheme. Therefore, the data size of the whole blockchain keeps growing. Fig.~\ref{img:btcchaingrows} shows the size of the Bitcoin blockchain. The size is rapidly increasing from the time that the system started in 2009, and in December 2018, the size is about 200GB. For this characteristic, to operate a Full Node, users need to prepare enough storage capacity. Moreover, users also need to add storage continually according to growing the blockchain. Therefore, users are difficult to operate a resource-constrained device (e.g., smartphone) as Full Nodes.

To address this issue, there should be a node architecture for a storage resource-constrained device to independently verify transactions so that like a Full Node. The architecture needs to reduce the required storage capacity for a node. There are some lightweight node architecture proposals such as SPV (Simple Payment Verification)~\cite{SPV}. The existing lightweight node architecture is that a lightweight node relies on Full Nodes. Therefore, existing lightweight nodes cannot verify transactions independently. In other words, the availability of existing lightweight nodes depends on the availability of Full Nodes it is relying on. Hence, without Full Nodes that lightweight nodes rely on, users of existing lightweight nodes cannot verify whether their transactions are processed normally in the Bitcoin network or not.

This thesis presents a storage load balancing scheme ``KARAKASA.'' In KARAKASA scheme, resource-constrained devices can verify transactions and blocks same as Full Nodes while reducing storage size requirement. We adopt a Distributed Hash Table(DHT)-based distributed storage to the storage of Bitcoin nodes. DHTs such as Chord~\cite{chord} realize effective node assignment for a block and lookup of Key-Value to peer-to-peer networked nodes. In KARAKASA, a subset of nodes in the Bitcoin network builds DHT clusters. Each node in clusters can behave as a Full Node; verify new transactions by referring others without keeping the whole blockchain.

By adopting securing methods for DHT~\cite{DHTsecuritysurvey}, compared with existing lightweight schemes, nodes in KARAKASA scheme keep the independent validation. Currently, a user should choice Full Nodes or node relying on Full Node. In other words, a user who has only resource-constrained devices needs to trust a Full Node that another user operates. By KARAKASA scheme, the user can operate a Bitcoin node without trusting another user.

\section{Organization of This Thesis} 
The thesis is organized as follows; In Chapter~\ref{background}, we explain Bitcoin and DHT as background materials. In Chapter~\ref{issue}, we discuss the issue of Bitcoin nodes. In Chapter~\ref{proposedscheme}, we present KARAKASA scheme and how to load balance the Bitcoin node storage using DHT. In Chapter~\ref{viewpointofevaluation}, we describe the evaluation plan of KARAKASA. In Chapter~\ref{Analysis}, we analyze KARAKASA from two point of views; storage capacity for a node in KARAKASA scheme and independent validation of transactions and blocks from security analysis of KARAKASA scheme. In Chapter~\ref{Discussion}, we compare KARAKASA scheme with Full Node and existing lightweight node scheme. In Chapter~\ref{related}, we introduce some works about existing lightweight nodes. In Chapter~\ref{conclusion}, we conclude this thesis and discuss possible future works.
\chapter{Background} 
\label{background} 
In this chapter, we describe the elemental technology of our work. First, we describe Bitcoin. Bitcoin is a first cryptocurrency based on the blockchain. Next, we describe Distributed Hash Table (DHT). DHT is a scheme for realizing efficient assignment and lookup in a peer-to-peer network.

\section{Bitcoin} 
This section describes how Bitcoin and Bitcoin nodes work. Users can operate Bitcoin nodes without depending on other nodes. It is considered a revolution in distributed systems.

\subsection{Bitcoin Overview} 
\label{BitcoinOverview} 
Bitcoin is a peer-to-peer payment system. Users show a payment of Bitcoin by creating and broadcasting data called ``transaction.'' The unit of Bitcoin is ``BTC\footnote{0.001 BTC is called 1 mBTC, 0.001 mBTC is called 1 $\mu$BTC, and 0.001 $\mu$BTC is called 1 satoshi.}.'' The inventor of Bitcoin focused on salving ``double-spending issue.'' Double-spending is like followings; An attacker pays a coin to another user. Then, the attacker pays the same coin to different users. To detect double spending, nodes store all transactions in a blockchain. Nodes store transactions as a unit of blocks. When a user pay Bitcoins, a node make a new transaction and broadcast to other nodes. When nodes receive a new transaction, nodes verify its correctness. Nodes verify with referring the local copy of the blockchain. Some of the nodes collect verified transactions and form a new block. Similar to transaction, a node broadcasts a new block. Other nodes receive it and verify. If the block is verified, nodes store the block in the local copy of the blockchain.

To show the order of blocks and of transactions included in blocks, Bitcoin uses chain of hash. A block includes a hash value of the previous block in the header. Additionally, the hash of the block is required lower than the predetermined difficulty value. Each node calculates the difficulty from per the last 2016 blocks in the blockchain. To create a block data that matches the requirement, a node needs to repeatedly calculate the hash value of the block with replacing data. The replaced data has no meaning. Once the node successfully creates a block that satisfies the requirement, the node broadcasts the block to the Bitcoin network. At this time, there is a possibility that another node broadcasts another block that refers to the same previous block. This situation is called a ``fork.'' When a fork happens, a node needs to choose one of contradicting blocks. In Bitcoin, a node selects a block so that the sum of the difficulty in local blockchain is the largest. There is an incentive mechanism to spend computation power to make a block. A node that makes a new block can get a reward as some Bitcoins. The reward is 12.5 BTC in December 2018. From a metaphor of gold mining, the block making process is called ``mining'' and the user working on mining is called ``miner.''

To withdraw transactions or to falsify, an attacker may try to rewrite a transaction. Then, the attacker rewrites a block that includes the transaction. Therefore, the hash value of the rewritten block would changes. The following block includes the hash value of the rewritten block. By the hash chain structure, hash values of all blocks following the rewritten block would change. Thus, the attacker needs to rewrite all block starting from the following block of rewritten block towards the latest one. It means that the attacker needs to rebuild another blockchain. At the same time, other honest nodes may create a new block that connects to the latest block in a existing blockchain. Hence, the attacker needs to make blocks more quickly, so that the attacker needs more computation power than honest ones. Whether a miner can make a block during the particular time is probabilistic. It depends on the computational power of the node apply. Hence, an attack success is probabilistic too. The inventor of Bitcoin showed attack success probability by random walk and Poisson distribution~\cite{bitcoin}.

\subsection{Bitcoin Transaction and Block Data Structure} 
\begin{figure}[h]
  \centering
  \includegraphics[width=14cm, bb=0 0 842 595]{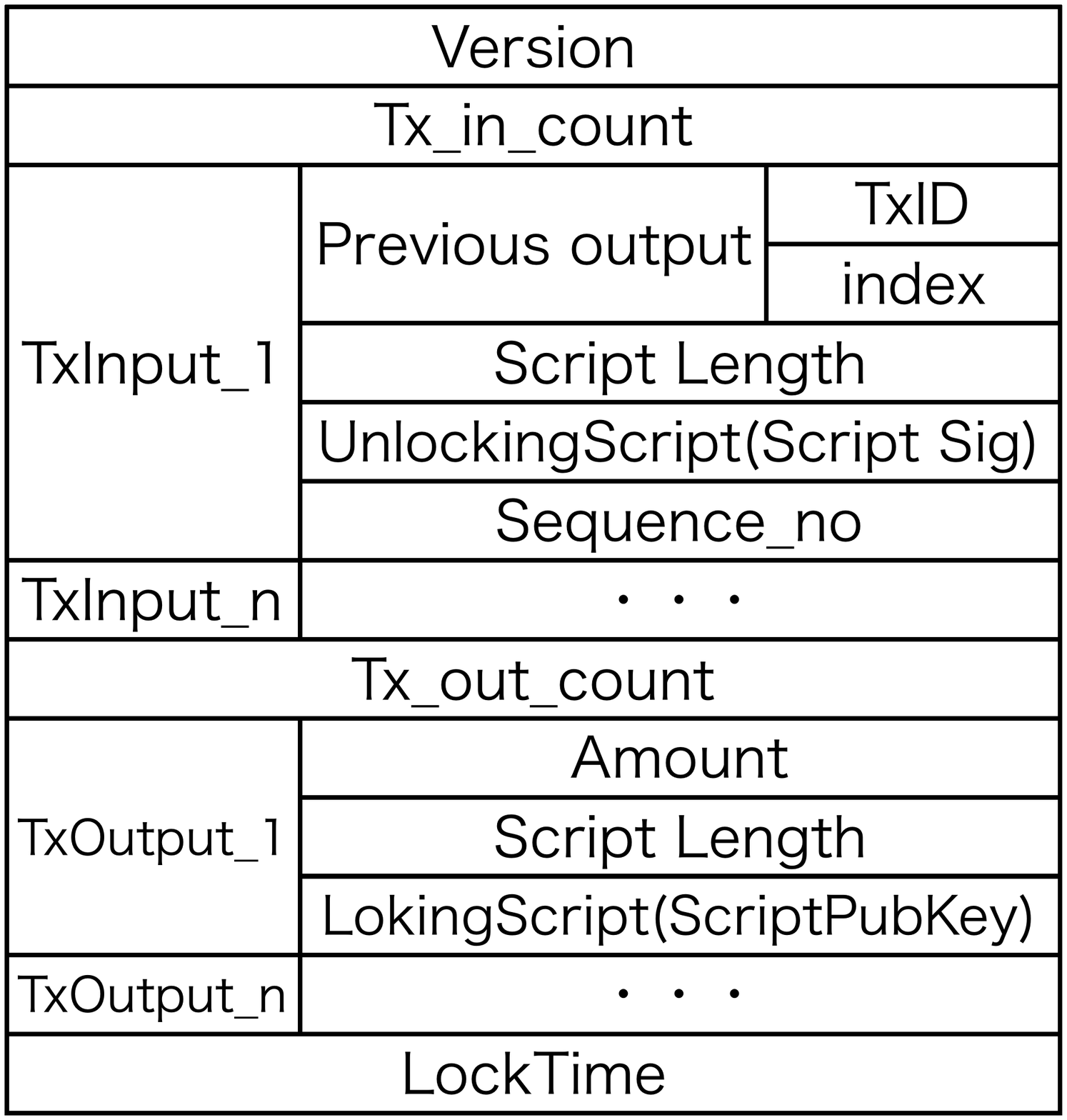}
  \caption{Bitcoin Transaction Data Structure}
  \label{img:txdata}
\end{figure}

This section shows Bitcoin transaction data and block data structure. Fig.~\ref{img:txdata} shows the data structure of a transaction~\cite{wikitx}. A transaction contains one or more TxInputs (TxIn) and TxOutputs (TxOut). In TxOut, {\it LockingScript } (it contains a template script {\it ScriptPubKey}) specify the requirement for paying coins. The amount of the coin also described in {\it Amount} field. A TxIn refers to a previous TxOut. {\it UnlockingScript} (it contains a template script {\it ScriptSig}) field in the TxIn shows that the creator of the transaction matches the requirement written in the referred TxOut. TxOuts that have not been referred by any TxIns in any other transactions are called ``Unspent Transaction Output (UTXO).''

\begin{figure}[h]
  \centering
  \includegraphics[width=14cm, bb=0 0 842 595]{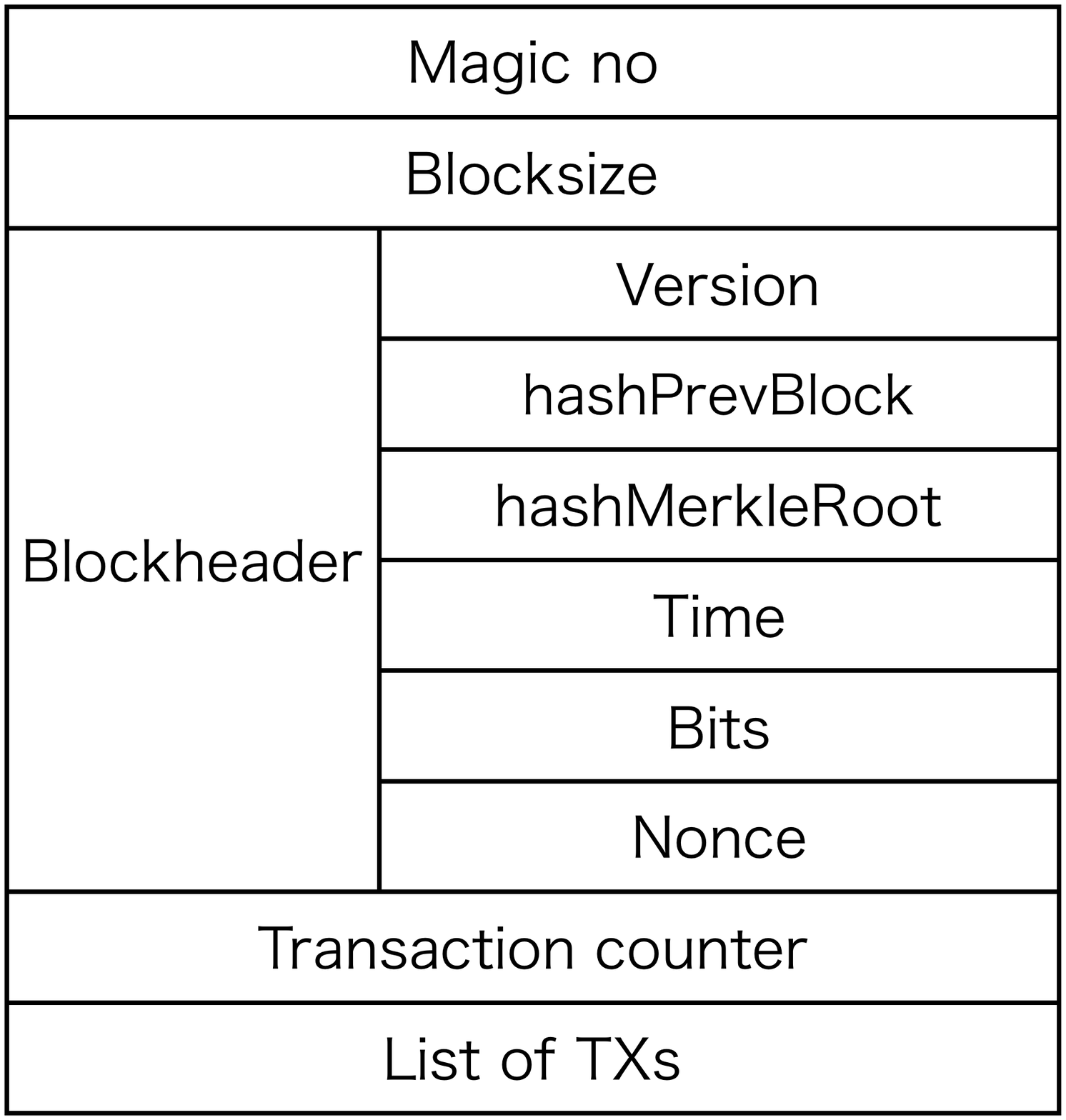}
  \caption{Bitcoin Block Data Structure}
  \label{img:blockdata}
\end{figure}

Fig.~\ref{img:blockdata} shows the block data structure. A block consists of block header and list of transactions. block header includes the followings; a hash value of the previous block {\it hashPrevBlock}), Markle root of transactions in the block ({\it hashMarkleRoot}), a target of Proof-of-Work ({\it Bits}), and a count of Proof-of-Work trial ({\it Nonce}). The {\it hashMarkleRoot} is a digest of the list of transactions in the block. Therefore, a hash of the block header can be taken as the digest of block. The {\it hashPrevBlock} is the hash of the block header of the previous block. The {\it hashPrevBlock} should satisfy the requirement described in Section~\ref{BitcoinOverview}.

\clearpage

\subsection{Bitcoin Transaction Verification and Node Storage} 
\label{VerifyTXandStorage} 
\begin{figure}[h]
  \centering
  \includegraphics[width=14cm, bb=0 0 842 595]{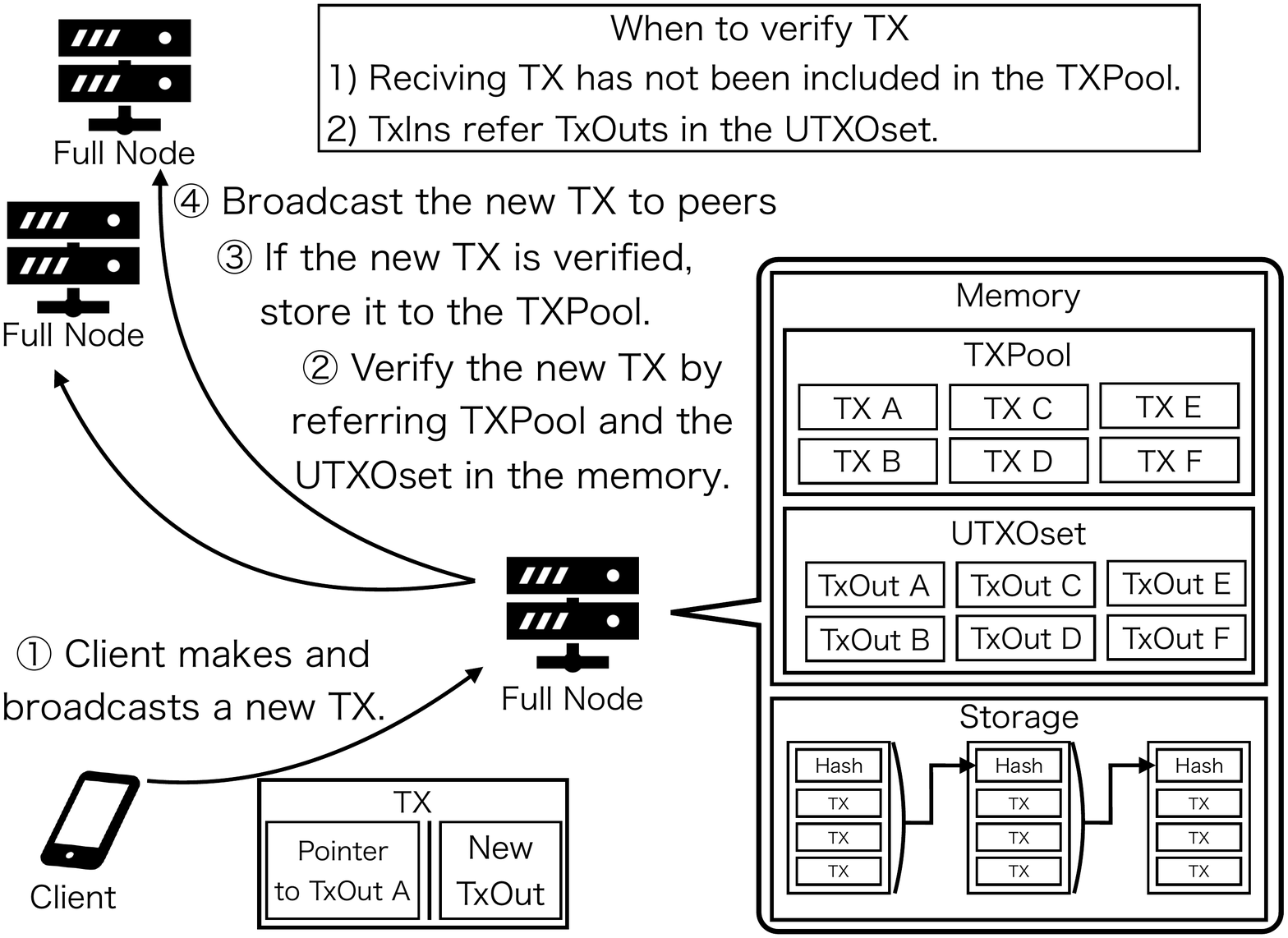}
  \caption{Verify Process and Node Storage of Bitcoin Transactions}
  \label{img:txverify}
\end{figure}

When each node receives a transaction, the nodes verify the transaction by the referring UTXOs. Fig.~\ref{img:txverify} shows Bitcoin transaction verification process and the relationship with node storages. To search UTXO in every verification process, a node needs to access the blockcahin in the local storage. To search efficiently UTXOs, in Bitcoin Core~\cite{Bitcoincore}, one of the implementations of Bitcoin, UTXO is pooled in a memory area. The memory area is called ``UTXOset.'' Nodes use it for transaction verification. Additionally, nodes store verified transactions in ``TXPool.'' Nodes create a block from transactions stored in TXPool.

\subsubsection{Simple Payment Verification(SPV)} 
The inventor of Bitcoin pointed out that Bitcoin is not scalable because of the append-only characteristics of the blockchain~\cite{bitcoin}. Hence, the inventor proposed a lightweight verification scheme. The scheme is called Simple Payment Verification (SPV). By SPV, a node processes the verification by referring information in active and on-line Full Nodes. Therefore, the node using SPV does not download the whole blockchain. A node using SPV holds only the block header chain instead of the whole blockchain. A node using SPV is called ``SPV node.''

SPV nodes can verify whether a new block is valid or not by checking the {\it hashPrevBlock}. It is the same verification process in Full Nodes. The difference is in the transaction verification process. Generally, A SPV node does not verify all transactions. The SPV node verifies transactions that describe payment to the user of the SPV node. A transaction in a blockchain is verified by a miner. Additionally, when other nodes accept a block include the transaction, the nodes also verify the transaction. Hence, a transaction in a blockchain can be considered as correct. The SPV node verifies a transaction by checking whether the transaction is included a block in a blockchain or not. When an SPV node verifies a transaction, the SPV node gets a Markle subtree from Full Nodes that the SPV node relies on. By using the Markle subtree, the SPV node can verify whether a transaction is included in a block or not. This SPV process does not need the whole blockchain. However, this process depends on whether the relied Full Nodes are honest.

\subsection{Bitcoin Node and Roles} 
\label{bitcoinnoderole} 
There are five Bitcoin node roles; holding blockchain, verifying new transaction and blocks, mining, P2P networking, and wallet. Each node has some or all role of those. Nodes that has all role is called ``Full Node.'' Ideally, all nodes should be Full Nodes. Then, nodes do not require to trust each other nodes. However, some roles need computation power or enough storage capacity. Thus, to use Bitcoin, resource-constrained nodes trust other nodes. They depend on the execution of some function on other nodes.

A node holding the entire of blockchain can detect and resolve forks. Forks happen by rewriting previous transactions or propagation delay of blocks. The probability of a fork is never 0. A node can create a new block that refers to a particular block that is already referred by another block created by other nodes. The new block can be accepted by other nodes with satisfying the selection rule described in Section~\ref{BitcoinOverview}. When a node observes a fork, the node needs to decide which is correct by referring to a previous block. Therefore, nodes need to keep all blocks. From another point of view, if a node leaves a network, other nodes holding the blockchain are not affected. They hold all data required to verify new transactions. On the other hand, an SPV node would not be available when the relied Full Node leaves. Hence, holding the whole blockchain contributes to the availability of the Bitcoin node.

Verifications of new transactions and blocks are stateless. Therefore, with keeping the whole blockchain, a node can verify and collect correct transactions and blocks independently from other nodes. Nodes need UTXOs for transactions verification. Nodes need the whole blockchain for blocks verification also.

Mining is a work for the confirmation of transactions also. By mining, A transaction is inserted in a block, and the block appended to the blockchain. Basically, rewriting the transaction in the blockchain is difficult according to honest computation power in the Bitcoin network and to how many blocks stacked on the block that includes the transaction. To work on mining, a node needs verified transactions and the last block in a blockchain.

P2P networking is used for propagating transactions and blocks. Nodes relay verified transactions and blocks. All nodes try to reach a consensus on only one blockchain in the network. There are several works focus on Bitcoin networks~\cite{hijackbtc2017, Decker2013, Apostolaki2018}. These works showed that it is necessary that enough number of Full Nodes share correct transactions and blocks.

Wallet function makes transactions for a payment. A wallet needs to recognize UTXOs that are associated with the wallet owner. The wallet also holds a cryptographic key pair. When the wallet makes a transaction, the wallet signs a new TxIn to express that the user is authorized to use a TxOut.

\subsection{Bootstrapping Full Nodes} 
\label{bootstrappingFullNode} 
This section describes the process of bootstrapping Full Nodes. To bootstrap a Full Node, the Full Node needs to discover peers in the Bitcoin network. After that, the Full Node gets the whole blockchain.

In Bitcoin Core, there are several schemes to discover nodes~\cite{disoverBTCpeer}. The primary schemes are followings; selecting IP addresses that an operator of the Full Node knows and sending a message of node discovering protocol to near hosts, or getting the IP address of Full Nodes from trusted parties. The third scheme is a default scheme in Bitcoin Core version 0.6 or later. The trusted party uses the DNS scheme~\cite{BootstrappingBTC}. The trusted party is called ``seed DNS.'' A bootstrapped Full Node resolves some domains. Then, the bootstrapped Full Node can get addresses of Bitcoin nodes associated with the domains. The domains are included in Bitcoin Core source code. If the bootstrapped Full Node fails to resolve domains, the bootstrapped Full Node try to connect directly addresses that are included in the source code.

The bootstrapped Full Node retrieves the whole blockchain from the connected Bitcoin nodes. The connected Bitcoin nodes send blocks. Since the first block is included in the Bitcoin Core source code, the bootstrapped Full Node can verify all block from the second block to the latest block. The bootstrapped Full Node verifies the received blocks. After receiving and verifying the latest block, the bootstrapped Full Node can behave independently.

The point of bootstrapping is that a Full Node verifies all block start from the first block to the latest block before acting as a active Full Node. When an operator wants to restart an inactive bitcoin node, the Bitcoin process also verifies the blocks in the storage. After verification, the Full Node receives blocks missing on the Full Nodes, up to the latest blocks on the blockchain.

\subsection{Cryptocurrency Exchange Service} 
In this decade, not only Bitcoin, many cryptocurrencies based on blockchain scheme were proposed. Those cryptocurrencies have different characteristic. One of the famous cryptocurrency is Ethereum~\cite{wood2014ethereum}. Ethereum is an application platform as well as a cryptocurrency. In Ethereum users can run programs called ``Smart Contract.''

\begin{figure}[h]
  \centering
  \includegraphics[width=14cm, bb=0 0 720 567]{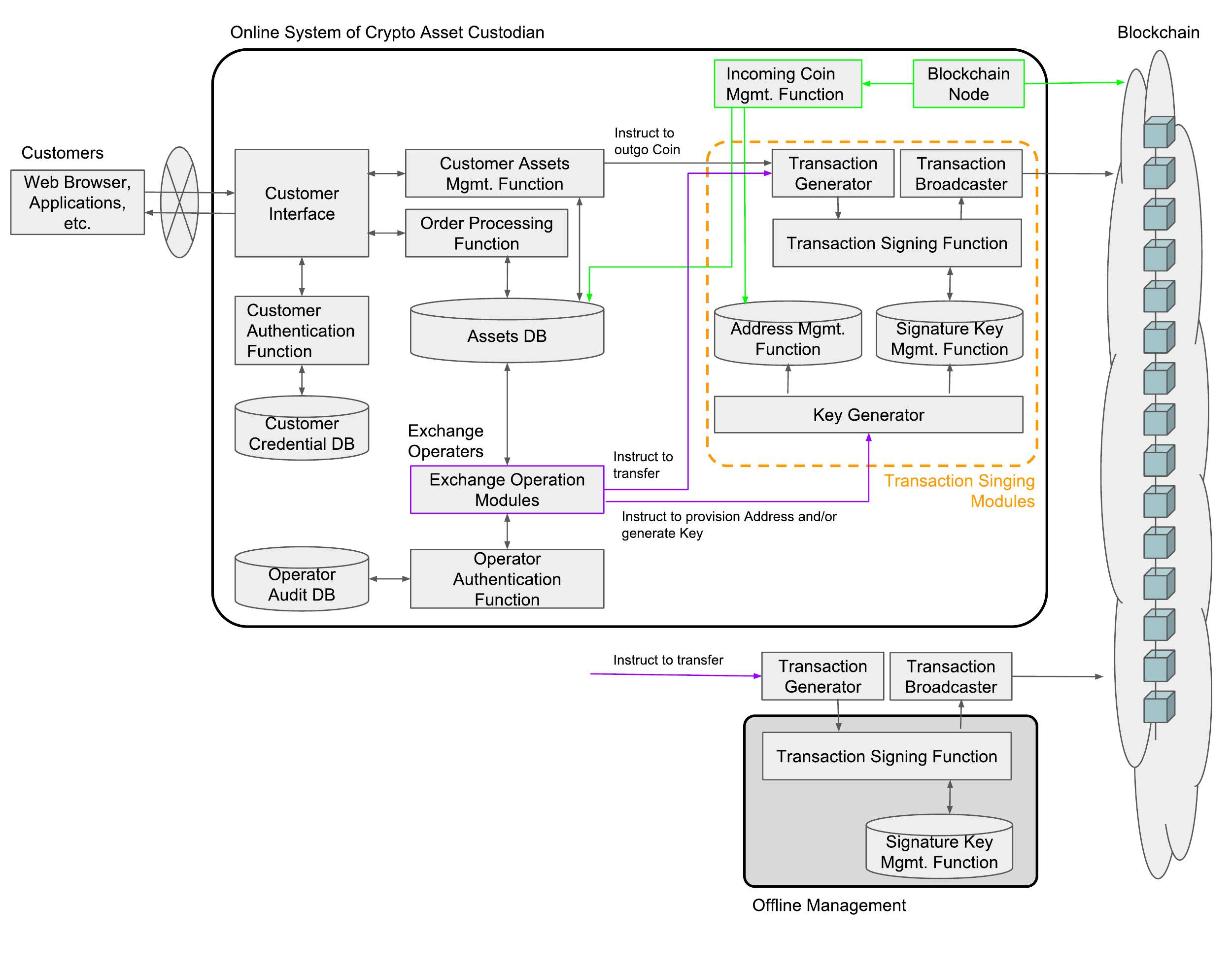}
  \caption{Architecture of Cryptocurrency Exchange Service cited from~\cite{VCGTF}}
  \label{img:vcgtf}
\end{figure}

Some cryptocurrency exchange services provide exchanging cryptocurrencies and fiat money like American Dollar. They also provide a cryptocurrency payment service. Those cryptocurrency exchange services are provided as web services and smartphone applications and so on. Standard architecture of cryptocurrency exchange services is discussed. Fig~\ref{img:vcgtf} shows standard service architecture that proposed in an Internet-Draft~\cite{VCGTF}. Generally, a service provider provides a server and operates the server as a Full Node. The client application requests processing payment to the server. Then, the server creates transactions and broadcasts it to a cryptocurrency network.

In 2018, there were several incidents that attackers hacked those services~\cite{coincheck, zaif}. The attacker maliciously sends cryptocurrencies instead of users. Those incidents were caused by hacking key management component and sign using user's secret key. In cryptocurrencies like Bitcoin, the ownership of coin is generally based on ownership of secret key. Hence, the architecture like Fig.~\ref{img:vcgtf} means that the security of user's cryptocurrency depends on the service provider.

From another aspect, when a user sends some cryptocurrencies another, the architecture lacks transparency for users. When a user requests sending, a Full Node in a service provider creates transaction and broadcasts it to a cryptocurrency network. At that time, because the client the user using is not a Full Node, the user cannot verify whether the transaction is created and broadcasted correctly or not. Bitcoin is designed as all nodes verify transactions and blocks independently. Hence, the cryptocurrency exchange service is not fitting to the design of Bitcoin.

\section{DHT (Distributed Hash Table)} 
\label{DHToverview} 
DHT is a design to implement a hash table among peer-to-peer networked nodes. Before Stoica et al. introduced a DHT implementation called ``Chord'', consistent hashing~\cite{consistenthashing} was used for key mapping in distributed systems. Consistent hashing is a design sharing a hash table among predetermined nodes. There was a problem that consistent hashing is not scalable with joining nodes. DHT is solving the problem by updating routing table dynamically. In DHT, each of the nodes store key mapping. It allows data location to be decided and be able to lookup in a distributed Key-Value-Store (KVS).

\begin{figure}[h]
  \centering
  \includegraphics[width=14cm, bb=0 0 842 595]{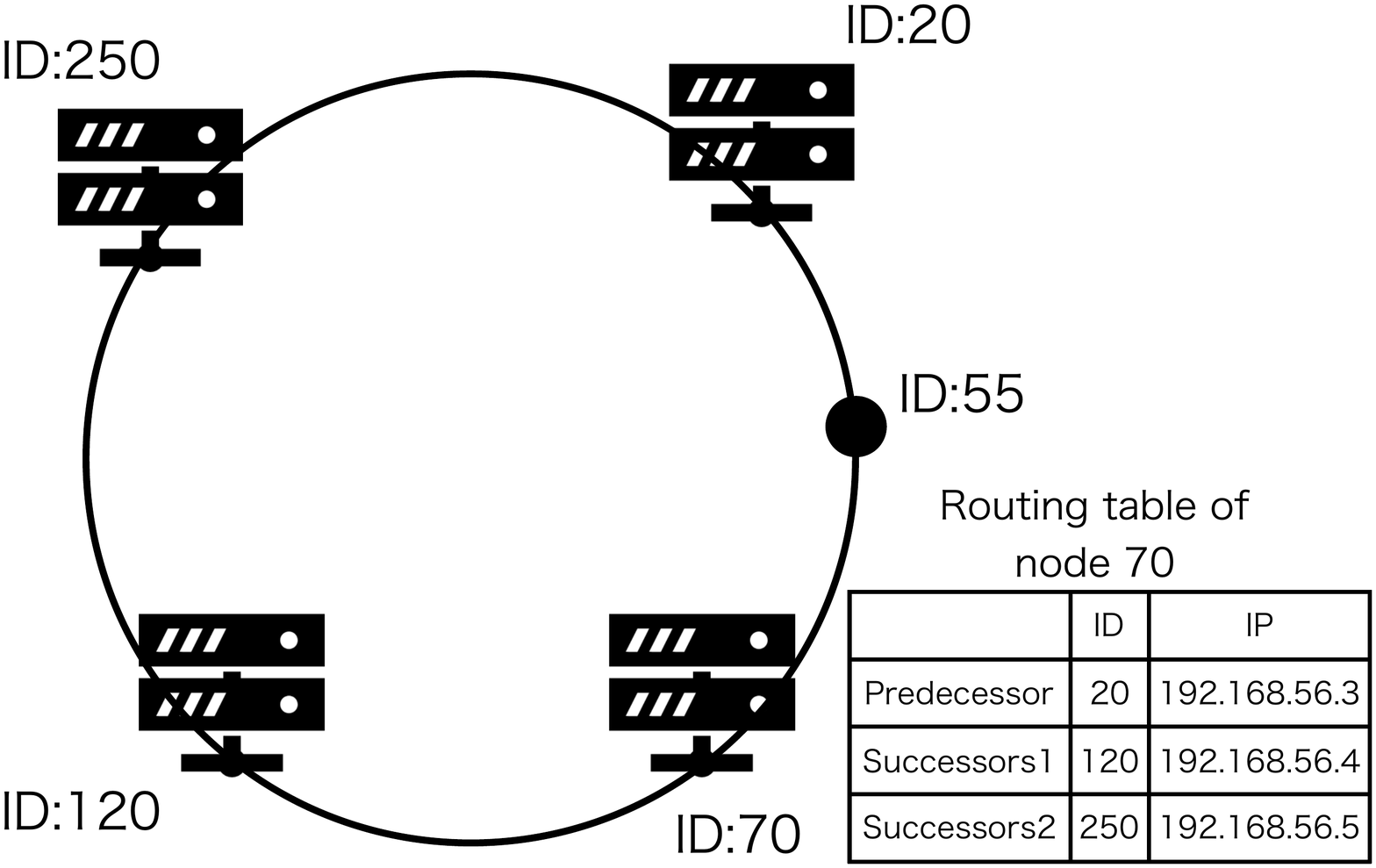}
  \caption{Chord overlay network and key mapping}
  \label{img:chord}
\end{figure}

Chord~\cite{chord}, one of the many DHT designs, uses a ring-shaped overlay network. Fig.~\ref{img:chord} shows an example of Chord overlay network and key mapping. The nodes and the data are mapped in a ring overlay network. A data ID is determined by calculating the hash value of data. A node ID is determined by the hash value of the node's IP address. A node uses a data IDs as the key for searching data from the distributed KVS. Nodes are responsible for returning data that are assigned IDs between an ID just after the previous node on the Chord ring and an ID of the node self. For example, in Fig.~\ref{img:chord}, the node ID 70 stores the data assigned to data ID 55. Each node has a routing table. The table includes next nodes in ring overlay (Successors) and previous (Predecessor). Additionally, the routing table includes routes to some nodes for efficient routing (Finger). By the routing table, when a network consists of $N$ nodes, key mapping lookup requires $\mathcal{O}(log N)$ messages. When nodes join or leave, each node updates the routing table. The process is called ``stabilizing.'' In Chord, updating the routing table requires only $\mathcal{O}(log^2 N)$. Chord provides an efficient scheme of calculating the location of data with few constraints on the applications~\cite{Dabekbp2ps2001}. Dabek et al. built a distributed file system called CFS (Cooperative File System)~\cite{Dabekwacs2001} based on their previous work based on Chord~\cite{Dabekbp2ps2001}.

For fault tolerance, there are some replication schemes proposed for Chord overlay. One of the schemes is neighbor replication. When a node has $N$ successors, $k$ $( k \leq N )$ successors have all data replicated in the nodes. For example, in Fig.~\ref{img:chord}, node ID 70 has two successors. When replicating two nodes, node ID 120 and 250 also have data ID 55. Then, when node ID 70 is down or leaves from the network, node ID 120 will be a successor of node ID 20. In addition, data ID 55 is still available by getting a replica from node ID 120.

There are DHT proposals other from Chord. CAN~\cite{CAN} uses a d-dimensional Cartesian coordinate space. Pastry~\cite{Pastry} uses Plaxton algorithm. Kademila~\cite{Kademlia} uses the XOR metric. Building CFS on top of any of these DHTs is possible.

\subsection{Security of DHT} 
\label{DHTsecurities} 
There are several discussions on the security of DHT~\cite{DHTsecuritysurvey}. The attacks for DHT are mainly three attacks; ``Sybil Attack'', ``Eclipse Attack'', ``Routing and Storage Attacks.''

\subsubsection{Sybil Attack} 
Sybil Attack is an attack where some malicious nodes join to a DHT cluster to achieve control of the DHT. Many nodes just joining is not obstruct processing the DHT protocol. However, this attack makes several attacks easier. Security of DHT is guaranteed under an estimation that $f$ of $n$ nodes in a DHT cluster are honest. An attacker breaks security by deploying virtual nodes over $f$ in a DHT cluster.

There are several countermeasures to Sybil Attacks. One of the countermeasures to Sybil Attack is deploying Certification Authority(CA). A CA authenticate when a node joins a DHT cluster~\cite{CAforsecureDHT, DCAforsecureDHT}. However, it needs that every node in the DHT cluster trusts the CA. Therefore, in this scheme, the CA is a single point failure of the DHT cluster. Another countermeasure is utilizing physical network characteristics~\cite{NetworkforSecureDHT, NetworkforSecureDHT2, NetworkforSecureDHT3}. Generally, an attacker deploys malicious nodes at close to attacker's other nodes in the network. utilizing this characteristics, a honest node can classify nodes by network characteristics. Additionally, honest nodes reject nodes joining if nodes deployed in nearby malicious nodes try to join. However, the countermeasure needs trusted network monitoring. From another aspect, a countermeasure using social networks has been proposed~\cite{SocialNetworkforSecureDHT}. Before joining a DHT cluster, operators share a key and nodes authenticate by the key. In this scheme, each node behaves like a small CA. It needs messaging outside of the DHT to exchange keys. Other countermeasures use computational puzzles and game theory~\cite{computationpuzzleforSecureDHT, computationpuzzleforSecureDHT2, gametheoryforSecureDHT}. These proposals work under an estimation that every node are rational. However, attackers process an attack because they can get profits (sometimes outside the DHT) by breaking the DHT. The rational countermeasure only on a DHT protocol needs discussion it is truly practical or not.

\subsubsection{Eclipse Attack} 
Eclipse Attack is an attack makes honest nodes can not lookup correctly by interrupting DHT lookup fowarding. Firstly, attackers control sufficient fraction of the neighbors of honest nodes. For example, the malicious nodes poisoning a routing table of victim nodes to establish control. Then, the malicious nodes deceiving the response of the routing table updating request. As a result, the victim node cannot receive correct responses of a lookup. Eclipse Attack is also effective to accelerate Routing and Storage Attacks. We would describe the Routing and Storage Attacks in the next section.

One of the countermeasures is to constrain how to assign the identities on the routing table. Chord uses this scheme~\cite{chord}. This scheme would work on conditions that node identities are random. Moreover, it is also required that malicious node spread over the DHT overlay. To spread malicious nodes, one of the schemes is that a CA assigns nodes random IDs. However, random assigning of identities has a trade-off with performance. In Pastry, there is optimizing mechanism of proximity neighbor selection~\cite{Pastry}. The mechanism utilizes weak requirements on the top level of the routing table. By adopting restrictions in the routing table, it is efficient that the network measurement and ID requirements for optimal neighbors selection. However, there is a problem that attackers can detect identities by exploiting the restrictions and the network measurement infrastructure.

Another countermeasure is redundant routing. In the condition that some node in a routing table are honest nodes, it will allow to keep redundant routes~\cite{redundantrouting}. Additionally, there is a proposal using a combination of an identity restriction and redundant routing~\cite{CAforsecureDHT}. This proposal utilizes that an attacker hijack the restricted routing table after hijacking optimized routing table. However, this scheme has overheads of using hijacked routing table and detecting a failure.

Countermeasures for Eclipse Attack have trade-offs between complexity and performance. Countermeasures need some works outside of DHT protocols. Therefore, to utilize them, it needs to consider overheads.

\subsubsection{Routing and Storage Attacks} 
Sybil Attack and Eclipse attack does not directly affect DHT. They accelerate attacks like refusing to forward a lookup request or responding incorrect values. Those attacks directly affect DHT. Those attacks are called ``Routing and Storage Attacks.'' Generally, an attacker attacks by forwarding to an incorrect, nonexisting or malicious node. For example, a combination of attacks described as follows; A node that relays a lookup request deliberately forward to a malicious node that joining by a Sybil Attack. Since it is easy to interrupt a routing, an attacker hijacks a routing table of a victim node by an Eclipse Attacks.

Main countermeasures of Routing and Storage Attacks are redundant routing and redundant storage. The Multi-Paths or the Multiple Wide Paths realizes redundant routing~\cite{CAforsecureDHT, redundantrouting}. By redundant routing, it guarantees reachability to honest nodes by keeping routes that does not include malicious nodes. Replications realize redundant storage. If a malicious node response incorrect value, nodes can check the correctness of the value by collecting replicas from several nodes.

There are three major replication schemes; replicating neighbor nodes in the ID space~\cite{CAforsecureDHT, redundantrouting, naiborraplica, naiborraplica2}, replicating random nodes in the ID space~\cite{randomreplica}, and the combination of them~\cite{combinationreplica, combinationreplica2}. The first scheme is easy to control the replication level and keeping consistency. However, it assumes that malicious nodes are spread over the ID space. If malicious nodes concentrate in a specific network region, a small number of malicious nodes can control replicas assigned on the network region. From this point of view, the second scheme is effective. Since the ID assignment scheme in DHT is public, it is easy to detect the network region assigned a specific key and attack them. Hence,  replication only is not enough as a countermeasure to storage attacks. It needs to be difficult to figure out where nodes are placed on the network from IDs.

The main concern to replication is whether replicas have the consistency or not. Whether each replica has been rewritten or not is verifiable on an application level. For example, if a key of a value is determined as a hash value of the value, the returned value is verifiable by checking the hash value of the response. The point of checking consistency is whether a node can verify independently or not. If a node lookup a key-value, and fails to verify key and response value, the node can choose a value by collecting replicas and metadata in the application level. There are several selection rules like choosing the highest version value, voting and so on.

\section{Summary of This Chapter}
In this chapter, we described the elemental technologies of this thesis, Bitcoin and DHT. Bitcoin is a peer-to-peer cryptocurrency system. By holding the whole blockchain, Full Node can verify new transactions independently. To process on lightweight devices, SPV is proposed. SPV nodes can verify new transactions with relying on Full Nodes. Cryptocurrency exchange services provide client applications and Full Nodes on their servers. DHT realizes efficient assignment and lookup among DHT networked nodes. You can implement distributed storage on top of DHT. There are some attacks to DHT; ``Sybil Attack'', ``Eclipse Attack'', ``Routing and Storage Attacks.'' To guarantee the security of DHT, several countermeasures have been proposed and discussed. Next chapter, we point out the issue we address in this thesis.
\chapter{Issues of The Thesis} 
\label{issue} 
There are several ways to use Bitcoin; using a Cryptocurrency Exchange Service, using an SPV node, or using a Full Node. This Chapter describes them and how the selection effects users. Then, We point out the issue address in this thesis.

verification
\section{Node Types and Verifications}
When using a Cryptocurrency Exchange Service, users use a client application for the service. Generally, described as in Fig~\ref{img:vcgtf}, the client application is not a Bitcoin node. A server of the service provider behaves as a Full Node. A client only sends a request of Bitcoin payment to the server. Sever would create and broadcast a transaction to Bitcoin Network. However, the client can not check whether the server processes the request or not. In the worst case, the service provider does not create a transaction. The service provider only updates a database that manages how much Bitcoin each user holds. Some service provider adopts this scheme because processing time of each transaction on the Bitcoin network is possibly more than ten minutes. In this case, the service provider should hold enough amount Bitcoin. The service provider will create a transaction only when a user withdraws Bitcoin outside of the service. In this scheme, there is no reason to use a cryptocurrency. It is just same as using fiat money and a bank. The availability for users depends on the server of the service provider. It means that the service provider is a single point failure.

When using an SPV node, as described in~\ref{VerifyTXandStorage}, users can verify a transaction is included in the Bitcoin blockchain. However, the SPV process depends on Full Nodes referred by the SPV node. Hence, the availability of the SPV node depends on the availability of Full Nodes the SPV node depends on. The SPV scheme is a scheme for resource-constrained devices. SPV nodes sometimes hold not all block headers of all blocks in the blockchain. SPV nodes hold only some number of block headers of latest blocks. SPV nodes cannot resolve all fork. They can resolve forks only if the forks start from a newer block that includes a block header the SPV node holds.

To verify transactions and blocks independently, and to resolve forks, a user needs to deploy a Full Node. By this scheme, a user can check independently whether a transaction is verified and included in a blockchain or not. To operate as Full Nodes, devices need enough size of storage for holding the whole blockchain.

\section{Issue of Storage} 
In the existing schemes, if a user wants to know independently whether a transaction is included in a blockchain or not, the user should operate a Full Node. When the user operates an SPV node, the user cannot process verification correctly if a Full Node that the SPV node depends on falsifies block headers and Markle trees. However, a Full Node needs enough storage capacity for storing the whole blockchain. On the other hand, users can not estimate enough storage capacity due to the nature of the append-only data structure of the blockchain. Hence, the user cannot operate a Full Node on a storage resource-constrained devices.

To address this issue, two requirements should be satisfied; One is to reduce the required storage size. Generally, the storage of a resouce-constrained device is built in and difficult to expand. Therefore, storage resource-constrained devices like smartphones cannot be append storage capacity. Additionally, from append-only characteristics, a blockchain would use up the storage eventually. Hence, to keep the node working, the storage needs to be scalable for holding the entire of the blockchain.

Another requirement is the independent verification transactions and blocks. The existing lightweight scheme SPV nodes rely on Full Nodes. As a result, the availability of an SPV node depends on Full Nodes. Additionally, SPV nodes cannot detect even if a Full Node maliciously sends a fake part of Markle tree. A malicious Full Node can send a fake transaction and a fake part of Markle tree as if the fake transactions is included in a blockchain. To detect fake transactions, a node needs to check independently with the whole blockchain.

\section{Summary of This Chapter}
In this chapter, we pointed out the issues of this thesis. When a user uses Bitcoin, a user can select options; operate a Full Node, operate an SPV node, use a client of a cryptocurrency exchange service. The latter two depend on other nodes. In other words, the user cannot independently verify whether his transaction is processed or not. To verify independently, user should operate a Full Node. However, it is difficult to keep operating a Full Node on a storage resource-constrained device because of the append-only characteristic of the blockchain. Hence, there are requirements for verifying transactions on a storage resource-constrained device; independent verification, and reduction of required storage capacity. In the next section, we propose new scheme for storage resource-constrained devices.
\chapter{Proposed Scheme: KARAKASA} 
\label{proposedscheme} 
\begin{figure}[h]
  \centering
  \includegraphics[width=14cm, bb=0 0 842 595]{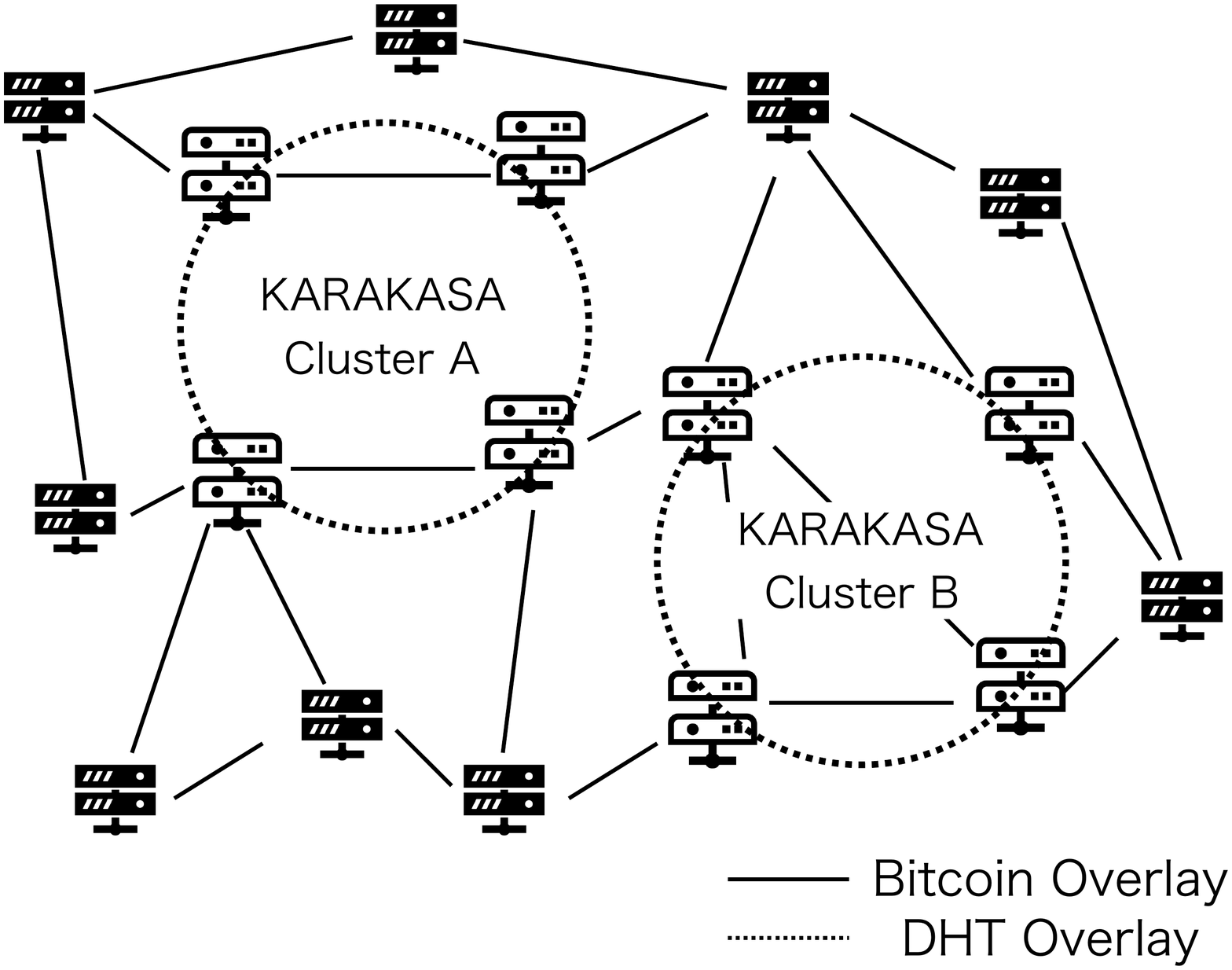}
  \caption{Sample Node Topology with KARAKASA Scheme; Some resource-constrained devices form DHT clusters. KARAKASA nodes can join the Bitcoin network the same as Full Nodes. Each KARAKASA nodes verify new blocks and new transactions by referring a blockchain distributed stored in the KARAKASA cluster. This figure is showing two KARAKASA clusters.}
  \label{img:topology}
\end{figure}

To address the issue described in Chapter~\ref{issue}, we propose a new scheme ``KARAKASA\footnote{The naming was inspired by Karakasa Renban Jo (\begin{CJK}{UTF8}{min}傘連判状\end{CJK}). Karakasa Renban Jo is a round robin scheme at middle ages (Shogun Era) in Japan. People who tried to revolution used this round-robin scheme. By arranging signatures in a ring, it made difficult to estimate who was a leader of the revolution. It was a load-balancing scheme of a responsibility.};'' load-sharing required storage capacity for each node by holding the whole blockchain among DHT clustered Bitcoin nodes. As Fig.~\ref{img:topology} shows, Users who operate storage resource-constrained devices collaborate their devices with each other and forms DHT clusters in a Bitcoin network. Each Bitcoin node in the cluster holds a part of blockchain assigned according to the DHT algorithm. When verifying new transactions and blocks, each node query the required block to the DHT cluster. Therefore, each node in the cluster can verify new transactions and blocks the same as Full Nodes. In this thesis, we discuss our scheme implemented on top of Chord. We call a DHT cluster of this scheme KARAKASA cluster, and call a node in a DHT cluster KARAKASA node.

\section{Node Architecture} 
Bitcoin nodes have two storage; ``BlockStorage'' stores actual block data, and ``ChainState'' stores UTXOs. To verify new transactions efficiently, nodes forms ``UTXOset'' in the node memory. We propose KARAKASA nodes keeps ``BlockStorage'' among DHT clustered nodes. When a KARAKASA node verifies new transactions, nodes only query to UTXOset. When blocks contradicting with a block in the blockchain are received, KARAKASA nodes query a contradicting block to the KARAKASA cluster to resolve the fork. This architecture makes nodes can behave same as Full Nodes in the verification process of transactions and blocks. By distributedly keeping the whole blockchain, the requirement for storage capacity for a node decrease.

\begin{figure}[h]
  \centering
  \includegraphics[width=14cm, bb=0 0 842 595]{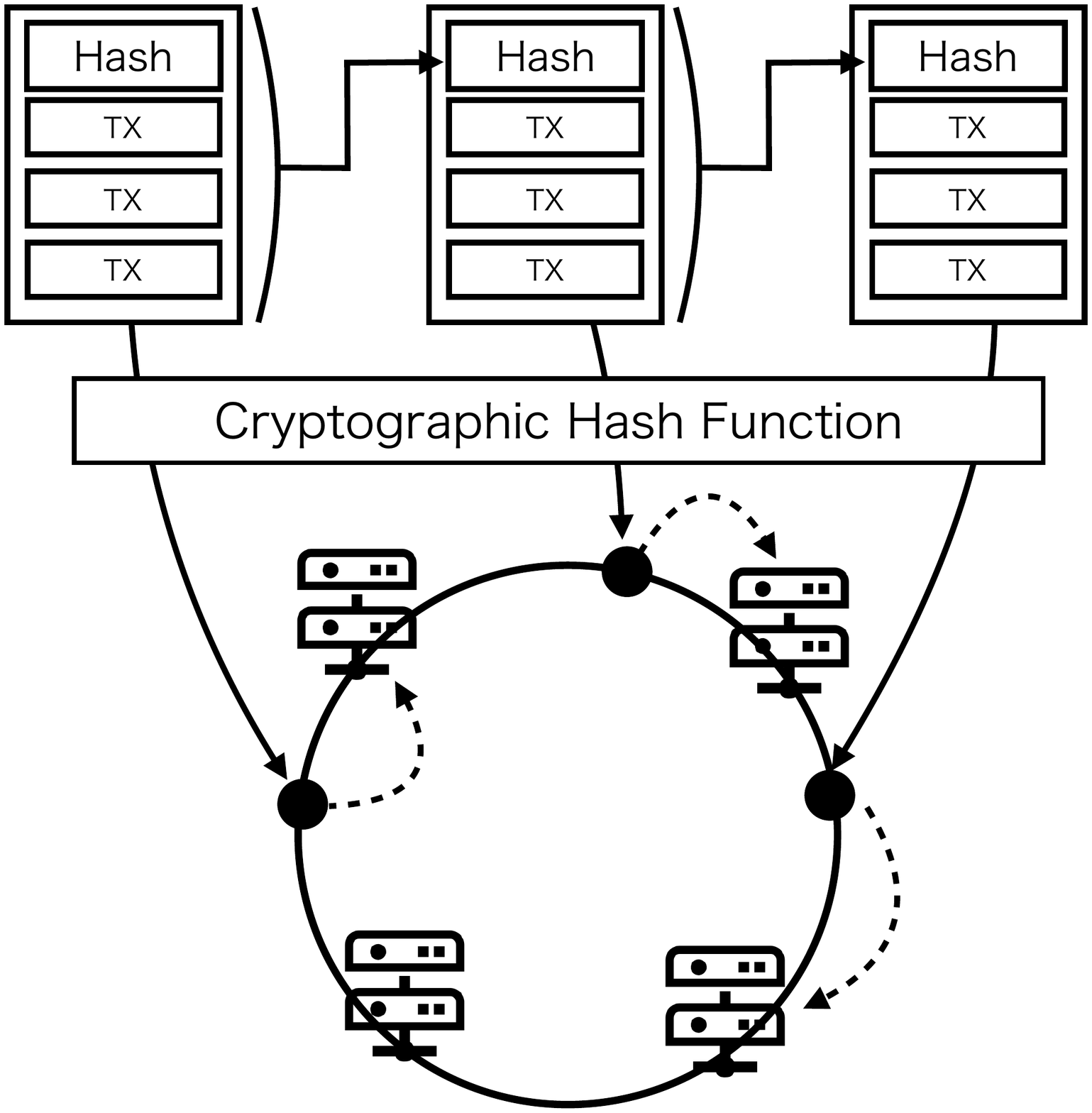}
  \caption{Blockchain on a Chord overlay network}
  \label{img:blockchainonchord}
\end{figure}

Next, we discuss how to map blocks in the blockchain to a Chord overlay network. Fig.~\ref{img:blockchainonchord} shows how blocks are mapped onto a Chord ring. To map blocks onto the Chord ring, the ID of each block in the Chord ring is determined by cryptographic hash function. Therefore, as Fig.~\ref{img:blockchainonchord} shows, the same KARAKASA node does not always store consecutive blocks. The replication guarantees fault tolerance.

\clearpage

It can be considered that a malicious KARAKASA node falsifies response for a block. In KARAKASA scheme, a node in the cluster lookups a block by an ID. The ID is calculated by a cryptographic hash function. Thus, honest nodes can detect rewriting by checking the ID and the hash of the response. When detect rewriting, the KARAKASA node can get the block from other nodes which replicate the block. Therefore, a malicious node needs to rewrite all of the replicas for successful rewrite. Therefore, replication schemes provide better tamper resistance.

\section{Bootstrapping and Scaling Process of a KARAKASA Cluster} 
\label{bootstrappingScalingDHT} 
When a node initializes a KARAKASA cluster, the bootstrapping process is like follows; First, create a DHT cluster consists of a single initial node. Second, the block retrieving process is the same as described in Section~\ref{bootstrappingFullNode}. In this block retrieving process, the initial node stores verified blocks in local. Since no other nodes have joined to the DHT cluster, storage in the initial node stores the whole blockchain. When another node joins the DHT cluster, the DHT stabilizing process assigns a part of the blockchain to the new node.

In the process that a node joins a DHT cluster, a node behave same as restarting Bitcoin process described in Section~\ref{bootstrappingFullNode}. Therefore, a new KARAKASA node needs to lookup all blocks from the KARAKASA cluster. Comparing with a Full Node, there is lookup overheads for building UTXOset. To achieve constant the lookup overheads, the new node needs enough time for the DHT stabilizing process to complete.

To secure a KARAKASA cluster, it needs countermeasure to attacks. A countermeasure to Sybil Attack is most needed. We adopt a scheme to guarantee the security of DHT described while Section~\ref{DHTsecurities} in bootstrapping and scaling process of a KARAKASA cluster. In our scheme, to refuse random hosts, an initial node configures cluster key. The initial node tells the key to other nodes. KARAKASA nodes authorize other nodes by the key in the join process. Using this scheme, the joining is restricted by whether the key shared or not. Curently, in this thesis, the key sharing depends on the operator's social network~\cite{SocialNetworkforSecureDHT}. Intoroduction of better key share shceme is out of scope of this thesis.

\section{Summary of This Chapter}
In this chapter, we proposed a new scheme ``KARAKASA'' to operate storage resource-constrained devices as like Full Node. In KARAKASA scheme, storage resource-constrained devices form a DHT cluster. The devices distributedly keep the whole blockchain among the DHT cluster. By DHT scheme, A KARAKASA node can retrieve required data in verification of transactions and blocks. Additionally, the replication scheme of DHT guarantees fault tolerance. In the bootstrapping and the scaling process of KARAKASA cluster, initial node configures a key to refuse that random hosts join. Those security schemes for DHT is available in KARAKASA scheme.
\chapter{Evaluation Plan} 
\label{viewpointofevaluation}
In this Chapter, we describe the point of views on the evaluation of KARAKASA scheme. KARAKASA scheme is a scheme for verifying transactions and blocks processing on a node while reducing required storage capacity for the node. Hence, we compare form two standpoints; First, as a performance evaluation, how much storage capacities are required and how overhead appear. Second, as a security point of view, whether a node can verify transactions and blocks like Full Node or not.

\section{Storage Capacity}
As described in Chapter~\ref{issue}, blockchain size would keep growing because of append-only characteristics of blockchain. Generally, storage of resource-constrained devices are built-in to the devices. Thus, it is typically difficult to keep adding enough storage to storage resource-constrained devices. Thus, a user cannot operate a storage resource constrained-devices as Full Node. 

KARAKASA scheme addresses the issue by storage load-balancing among DHT networked nodes. Thus, we evaluate the required amount of storage for a node. Moreover, we also analyze overhead for the distributed store of the whole blockchain in Bitcoin scenario. By reduction of the required storage capacity for a node, storage resource-constrained devices may store the verification of transactions and blocks on its local storage. 

\section{Independent Verification of Transactions and Blocks}
Originally, in Bitcoin design, the verification process of transactions and blocks are independent with other nodes. As described in Section~\ref{VerifyTXandStorage}, Full Nodes can verify independently with referring the local storage. To process a Bitcoin node on a storage resource-constrained device, SPV node needs to trust Full Nodes that it refers. However, when the Full Node that an SPV node rely on is dishonest, the SPV node verify transactions as correct even though transactions are in-correct. In other words, to reduce the required storage capacity, the SPV nodes does not keep the independent verification of transactions and blocks.

KARAKASA scheme aims to achieve both the reduction of required storage capacity and the independent verification of transactions and blocks. To achieve them, KARAKASA scheme utilizes characteristics of DHT. Thus, whether a KARAKASA node is independent or not depends on whether a DHT networked node is independent or not. Originally, in design scheme of DHT, DHT networked nodes can behave independently from other nodes. Thus, KARAKASA nodes can be considered that KARAKASA nodes can behave independently. The several attacks described in Section~\ref{DHTsecurities} obstruct DHT process so that nodes can not behave independently. Therefore, we considered that the independent verification of KARAKASA nodes is affected by the security of DHT. For example, if the node cannot retrieve a block from a KARAKASA cluster, a KARAKASA node cannot verify transactions and blocks. Thus, We analyze the security of DHT in Bitcoin scenario as the analysis of the independent verification.

\section{Summary of This Chapter}
In this chapter, we described the evaluation plan of KARAKASA scheme. The points are storage capacity for a node and independent verification of transaction and blocks. KARAKASA scheme aims to achieve both of the reduction of required storage capacity and the independent verification of transactions and blocks. To achieve them, KARAKASA scheme utilizes characteristics of DHT. Thus, whether a KARAKASA node is independent or not depends on whether a DHT networked node is independent or not. Therefore, we analyze the security of DHT in Bitcoin scenario as the analysis of the independent verification on KARAKASA nodes. In the next chapter, we analyze the two aspects of KARAKASA scheme.
\chapter{Analysis of KARAKASA}
\label{Analysis}
In this chapter, we analyze KARAKASA scheme from two aspects described in Chapter~\ref{viewpointofevaluation}. Firstly, we analyze KARAKASA scheme in terms of storage capacity, and messaging overhead among KARAKASA nodes for performance analysis. Secondly, we analyze the independent verification of transactions and blocks on a KARAKASA node. For analyzing the independent verification, we discuss the security of DHT in the Bitcoin scenario.

\section{Performance Analysis}
In this section, we analyze the performance of KARAKASA scheme. We analyzed the performance of KARAKASA scheme in the two points of view; storage capacity in a node and messaging overhead among KARAKASA nodes. We estimated common simulation parameters, then simulated KARAKASA scheme by Overlay Weaver~\cite{overlayweaver}. Table~\ref{table:variable} shows the common simulation parameters. In Bitcoin, the largest block size is 1MB. Thus, we defined $BlockSize$ as 1MB. $N$ is the number of nodes in the simulated network. $Suc$ shows the number of successors in a routing table that a node refers. $R$ specify the number of replicas created.

\begin{table*}[ht]
  \renewcommand{\arraystretch}{1.3}
  \caption{Simulation Parameters}
  \label{table:variable}
  \centering
  \begin{tabular}{|c|c|c|}
    \hline
    Symbol & Restriction & Description\\
    \hline
    \hline
    $BlockSize$ & $BlockSize = 1MB$ & \begin{tabular}{c} Size of a Block\\in bytes \end{tabular}\\
    \hline
    $BlockCount$ & $BlockCount \geq 1$ & \begin{tabular}{c} Number of Blocks\\in a Bitcoin blockchain\end{tabular}\\
    \hline
    \hline
    $N$ & none & \begin{tabular}{c} Number of Node\\in a cluster\end{tabular}\\
    \hline
    $Suc$ & $N-1 \geq Suc $ & \begin{tabular}{c} Number of Successors a node has\end{tabular}\\
    \hline
    $R$ & $(Suc \geq R)$ & \begin{tabular}{c} Number of replicas\end{tabular}\\
    \hline
  \end{tabular}
\end{table*}

\subsection{Storage Capacity}
\label{StorageSizeForANode}
In this section, we estimated storage capacity for a KARAKASA node. We verify the estimation by simulation. In the simulation, we adopt the parameter in the Bitcoin network. 

\subsubsection{Estimation}

We estimated the required storage capacity in a KARAKASA node. When a Full Node store the whole blockchain, the storage capacity of the Full Node is
\begin{eqnarray}
Storage Capacity_{FullNode} \approx BlockCount \cdot BlockSize
\end{eqnarray}

In our proposal, each KARAKASA nodes keeps a part of the blockchain. As we described in Chapter~\ref{proposedscheme}, the ID of a block is the result of a cryptographic hash function. The results of a cryptographic hash function consist of a uniform distribution. Therefore, in KARAKASA scheme, a number of blocks that a node holds also consist of uniform distribution. Without considering replication, the storage capacity of a KARAKASA node is as follows:

\begin{eqnarray}
\label{eq:nowcost}
Storage Capacity_{\mathit{KARAKASANode}} \approx \frac{BlockCount \cdot BlockSize}{N}
\end{eqnarray}

With KARAKASA scheme, replication is required for fault tolerance. For the evaluation for replication, we simulated replication scheme described in Section~\ref{DHToverview}. Storage capacity depends on numbers of replicas created. The storage capacity of the KARAKASA node can be calculated as follows;
\begin{eqnarray}
\label{eq:proposalcost}
& &Storage Capacity_{\mathit{KARAKASANodeWithReplication}} \nonumber \\
\nonumber \\
&\approx& \frac{BlockCount \cdot BlockSize}{N} + \frac{BlockCount \cdot BlockSize}{N} \cdot R \nonumber\\
\nonumber \\
&=& \frac{BlockCount \cdot BlockSize}{N} \cdot (R+1)
\end{eqnarray}
\noindent
A formula of estimating storage amount is just a simple linear equation to the number of nodes ($N$). When KARAKASA nodes replicate all blocks, storage amount is the same as Full Nodes. This situation means every KARAKASA node keeps the whole blockchain.

\subsubsection{Experiment}
\begin{figure}[h]
  \centering
  \includegraphics[width=14cm, bb=0 0 576 432]{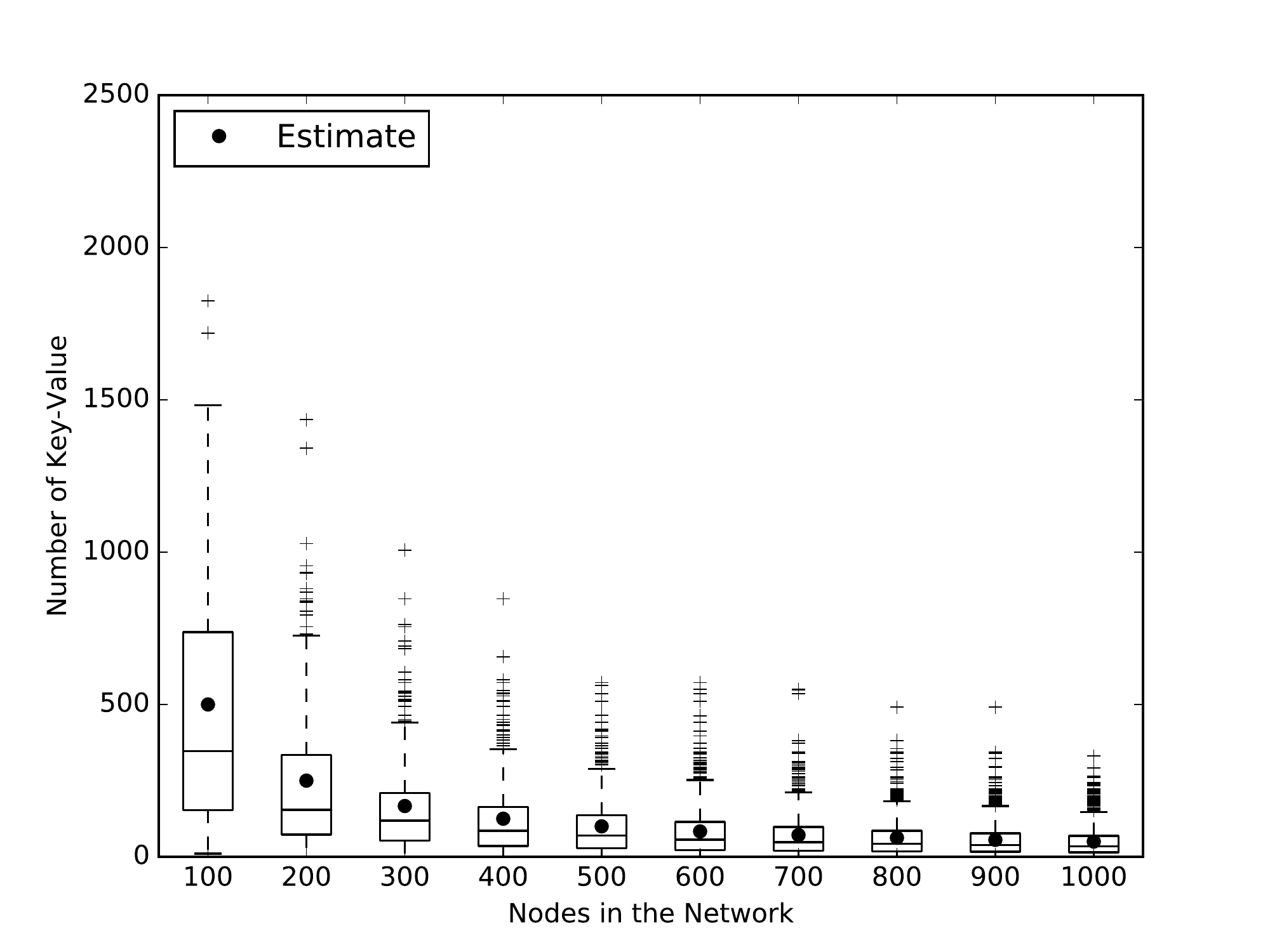}
  \caption{Node Number vs storage Capacity: This figure shows simulation result closed to estimated values.}
  \label{img:costvsnode}
\end{figure}

To verify our estimation, we selected simulation parameters from the current Bitcoin network. In March 2018, the number of nodes in the Bitcoin network is about 11500~\cite{bitcoinnodes}, and highest block number is about 512000~\cite{bitcoinblocks}. Thus, We simulated the storage capacity when storing 512000 blocks ($BlockCount$) with active number of nodes in 500 nodes to 1000 nodes in 100 intervals ($N$). Fig.~\ref{img:costvsnode} shows the simulation result. As expected in Formula~\ref{eq:nowcost}, more nodes in the KARAKASA cluster, fewer blocks are stored in one node. Therefore, more nodes in the KARAKASA cluster, the smaller storage capacity required.

\begin{figure}[t]
  \centering
  \includegraphics[width=14cm, bb=0 0 576 432]{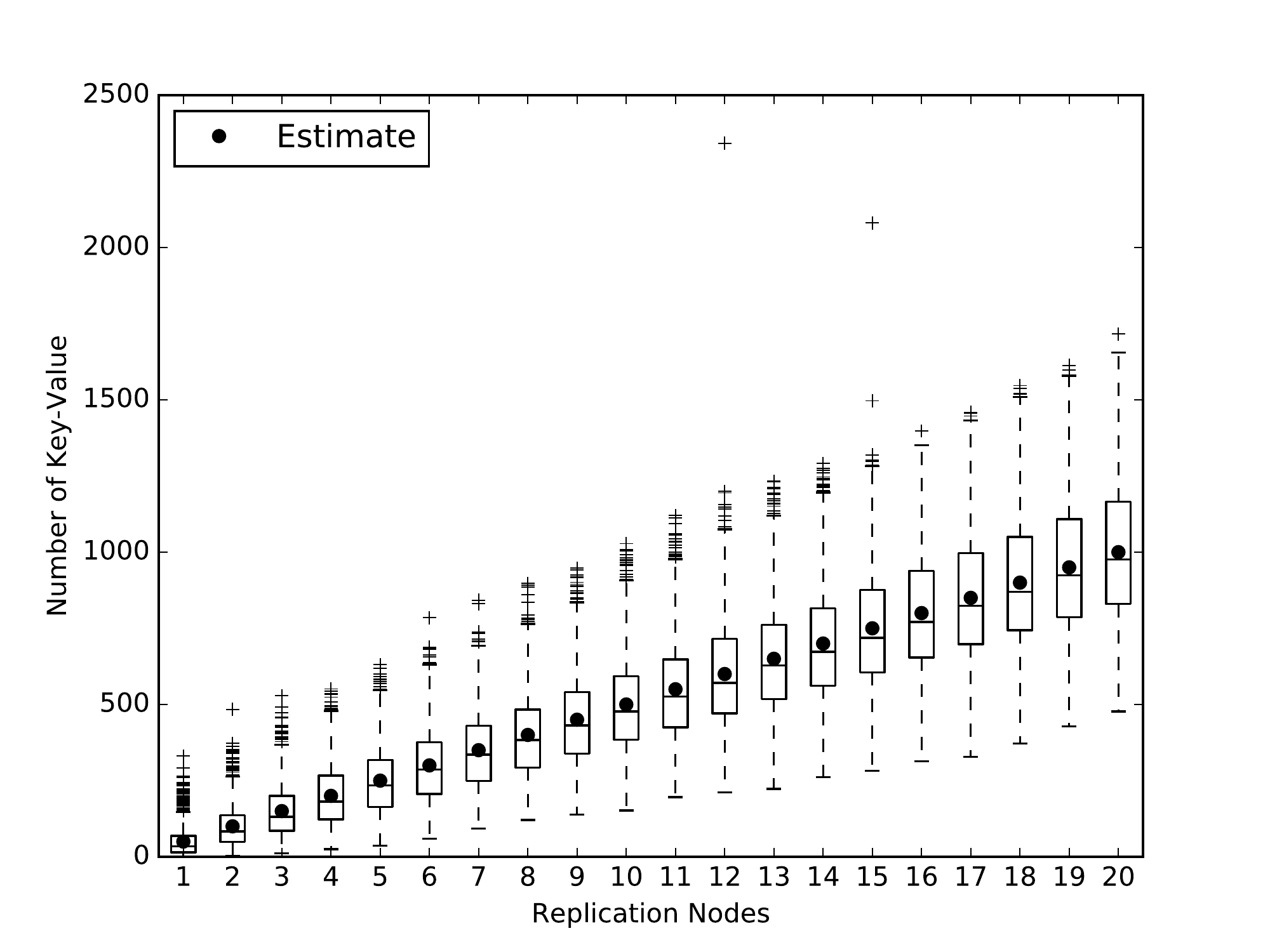}
  \caption{Replication vs Storage Capacity}
  \label{img:replicationvscost}
\end{figure}

Then, we simulated the storage capacity with replication. We set the number of nodes in a KARAKASA cluster ($N$) to 1000, and the number of blocks ($BlockCount$) to 50000. Fig.~\ref{img:replicationvscost} shows the simulation result of the relationship between replication node number and storage capacity. It shows that the more replication nodes are, the more storage capacity is needed. It is nearby with our estimation (Formula~\ref{eq:proposalcost}).
\clearpage

\subsection{Messaging Overhead}
\label{MessagingOverhead}
We analyzed messaging overhead for querying the DHT in Bitcoin scenario.

\subsubsection{Estimation}
In our proposal, each of the KARAKASA nodes hold a part of ``BlockStorage.'' On the other hand, the UTXOset is not distributed. Hence, when nodes verify a new block or new transactions, nodes do not query the KARAKASA cluster. When nodes resolve a fork of the blockchain, nodes query the starting block of the fork. Thus, if using Chord, the number of messages as overhead is as follows:
\begin{eqnarray}
OverheadForReadingABlock \approx \mathcal{O}(log N)
\end{eqnarray}

When a node builds UTXOset from the whole blockchain, the node needs to read all blocks in a blockchain. This process happens in joining process to KARAKASA cluster. Then, the number of the message to build UTXOset is as follows;
\begin{eqnarray}
\label{eq:overheadbuildUTXOset}
OverheadforbuildingUTXOset \approx BlockCount * \mathcal{O}(log N) 
\end{eqnarray}
This formula is a linear equation to $BlockCount$. Hence, as blockchain grows, the cost will be higher. There is a need to consider the messaging overhead.

\subsubsection{Experiment}
\begin{figure}[h]
  \centering
  \includegraphics[width=14cm, bb=0 0 576 432]{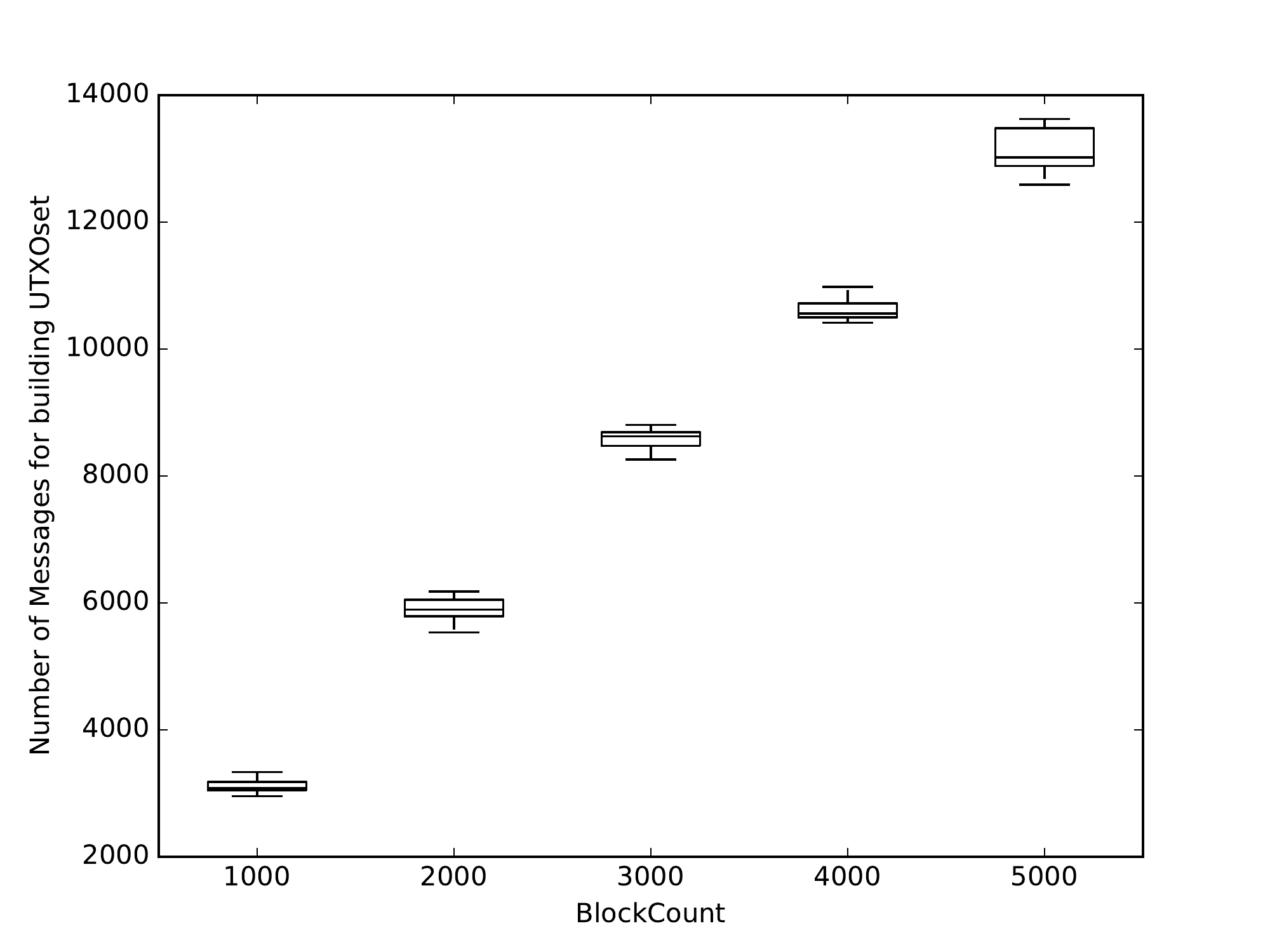}
  \caption{Message Overhead for building UTXOset} 
  \label{img:buildUTXOset}
\end{figure}

We simulated reading all blocks when the number of nodes ($N$) is 1000 and the number of blocks ($BlockCount$) in 1000 to 5000 in 1000 interval. Fig.~\ref{img:buildUTXOset} shows the results for each $BlockCount$ simulated ten times each. It shows that message overhead looks like a linear equation to $BlockCount$, same as our estimation (Formula~\ref{eq:overheadbuildUTXOset}).

\section{Analysis of Independent Verification of Transactions and Blocks}
\label{Security}
To achieve the independent verification of transactions and Blocks, nodes need to access historical transactions and blocks correctly. KARAKASA scheme utilizes DHT scheme to distribute data among nodes. Thus, the independent verification is affected by the security of storage based on DHT.

As described in Section~\ref{DHTsecurities}, main attacks for DHT are Sybil Attack, Eclipse Attack, Routing and Storage Attacks. The KARAKASA scheme utilizes social network countermeasures for Sybil Attack as described in Section~\ref{bootstrappingScalingDHT}. Like that, some countermeasures described in section~\ref{DHTsecurities} for eclipse attacks are available. In this section, we discuss the KARAKASA scheme on Chord. Chord has some stabilizing process for churns~\cite{ChordinChurn}.

In the Bitcoin context, the main concern is the transaction rewriting attack by utilizing Routing and Storage Attacks. We describe how to manage the transaction rewriting attack in a KARAKASA scheme and effects.

\subsection{Transaction Rewriting Attack}
\label{TXrewriting}
One of the attacks is rewriting a particular transaction in a block.  Thus, when an attacker rewrites transactions, a hash value of the block that includes the transactions would change. A hash value of the block determines the ID of a block on KARAKASA cluster. Therefore, by checking hash values of response and the ID used for retrieving, a KARAKASA node can detect whether the response is faked or not. Hence, just rewriting a transaction does not work effectively. Additionally, if rewriting is detected, a node that lookup a block can get replications. Even if some of the replicas are also faked, nodes can take the correct replica so that the ID used for retrieving is same as the hash value of replica.

To succeed the rewriting attack, an attacker needs to access all KARAKASA nodes that keep a block including a target transaction. The KARAKASA nodes keep a part of the whole blockchain. Thus, the attacker needs to search a block include the target transaction first. Moreover, the attacker needs to search a node keeping the block. Those requirements are overheads of the attack comparing to attacking a single Full Node. As described in the previous paragraph, the attacker needs to rewrite all replications. Hence, the attacker also needs to lookup nodes that replicate the block.

By the hash chain structure of blockchain, If a block in a blockchain is rewritten, the following block also changes. Therefore, an attack does not succeed only by rewriting a block including a target transaction. The attacker needs to rewrite all blocks stacked on the rewritten block. The generalized situation is following; An attacker wants to rewrite a transaction. $N$ blocks are stacked on the block that includes the target transaction. A KARAKASA cluster creates $Replica$ number of replicas. Due to this design, the attacker needs to rewrite $Rplica \cdot N$ blocks. It means that more blocks are stacked on a block, an attacker needs to rewrite more data. In other words, the security parameter is count of the replica times count of stacked blocks. As described in Section~\ref{StorageSizeForANode}, more replication needs more storage capacity. Thus, there is a trade-off between storage capacity and security.

\subsection{Effect of Transaction Rewriting Attack}
Even if an attacker succeeded in a transaction rewriting attack, it means just that some nodes in the Bitcoin network accept rewritten blocks. In the proposed scheme, a KARAKASA cluster is a part of the Bitcoin network. Hence, even if an attack to a KARAKASA cluster is a success, the rewritten transaction is not accepted by other nodes. To be accepted to other nodes in the Bitcoin network, the block that includes the rewritten transaction needs to be accepted. In other words, it needs that all nodes take the block as a correct block. It requires spending computation power because of Proof-of-Work consensus scheme. Hence, just rewriting a transaction is not enough as an attack.

On the other hand, in the double-spending attack, the rewriting attack has some effects. The author of this thesis showed in the previous work that the model of the double-spending attack and its incentive model~\cite{mercaripaper}. In the process of the double-spending attack, the attacker needs that a victim cannot detect a payment that is contradicting to payment to the victim. The propagation delay of transactions and blocks and the storage rewriting attack create this situation. If rewriting storage of the victim's node, the victim verifies a transaction correctly even though other nodes fail to verify. By communicating with other nodes and the fork resolving scheme, the victim would notice that verification is wrong. In this thesis, the incentive model of double-spending attack is out of scope. However, that analysis on KARAKASA scheme is necessary. 

\subsection{Conclusion of Analysis of Independent Verification}
By utilizing characteristics and security advantage of DHT with replication, the transaction rewriting attack does not work. Thus, storage based on DHT can be considered as secure even though distributedly store the whole blockchain. In other words, a KARAKASA node does not need to trust other nodes in the same KARAKASA cluster. Additionally, the KARAKASA node also does not need to trust nodes outside the KARAKASA cluster. The KARAKASA node can verify new transactions and blocks with referring the whole blockchain in the KARAKASA cluster. Hence, we conclude KARAKASA nodes keep the independent verification of transactions and Blocks.

\section{Summary of This Chapter}
In this chapter, we analyzed two aspects of KARAKASA scheme; performance and the independent verification. As a performance analysis, we estimate and simulate KARAKASA scheme. As a result, a required storage capacity for a KARAKASA node is reduced compared with Full Node. On the other hand, messaging overhead is like a linear equation to count of blocks in a blockchain. As an analysis of the independent verification, we discussed the security of DHT in Bitcoin scenario. Especially, we concluded that a transaction rewriting attack does not work on KARAKASA cluster. As a result, KARAKASA scheme satisfies to reduce required storage capacity for a node and to keep the independent verification of transactions and blocks.
\chapter{Discussion}
\label{Discussion}
In this chapter, we evaluate the KARAKASA scheme by compare with Full Nodes, and SPV nodes. KARAKASA scheme is a scheme for verification of transactions and blocks processing on the local of a node for reducing a required storage amount for one node. Hence, we compare form two standpoints; First, how much storage capacities needed. Second, whether a node of each scheme can verify transactions and blocks independently or not. Table~\ref{table:evaluation} shows the summary of the characteristics of each scheme.

\begin{table}[ht]
  \renewcommand{\arraystretch}{1.3}
  \caption{Summery of Characteristics}
  \label{table:evaluation}
  \centering
  \begin{tabular}{|c|l|l|}
    \hline
    Node Type & Storage Capacity & Verification of TXs and Blocks \\
    \hline
    \hline
    Full Node & \begin{tabular}{l} \textperiodcentered Full Blockchain \end{tabular} & \begin{tabular}{l} \textperiodcentered Independently \& locally \end{tabular} \\
    \hline
    \begin{tabular}{c} KARAKASA\\Node \end{tabular} & \begin{tabular}{l} \textperiodcentered A part of Blockchain \\ \textperiodcentered Scalable with scaling\\node number in the cluster \end{tabular} & \begin{tabular}{l} \textperiodcentered Independently on new TXs \\ \textperiodcentered Resolving Fork depends on\\ Replication count \\ in a KARAKASA cluster \end{tabular} \\
    \hline
    SPV Node & \begin{tabular}{l} \textperiodcentered Blockheaders \end{tabular} & \begin{tabular}{l} \textperiodcentered Depends on a Full Node \end{tabular} \\
    \hline
  \end{tabular}
\end{table}

\section{Comparison with Existing Schemes}
In this section, we compare KARAKASA node, Full Node, and SPV node. With referring analysis in Chapter~\ref{Analysis}, we point out some advantages of KARAKASA scheme.

\subsection{Storage Capacity}
We describe the estimation of the required storage capacity for Full Node, SPV and KARAKASA. In Section~\ref{StorageSizeForANode}, the storage capacity required for Full Node and KARAKASA node was shown. In this section, we also describe and compare with the storage capacity of an SPV node.

\begin{table}[h]
  \renewcommand{\arraystretch}{1.3}
  \caption{Storage Parameters}
  \label{table:StorageParameters}
  \centering
  \begin{tabular}{|c|c|c|c|}
    \hline
    Symbol & Unit & Description & Constant or Variable\\
    \hline
    \hline
    $BlockSize$ & MB & \begin{tabular}{c} Data size\\of a Block \end{tabular} &  1 MB Maximum \\
    \hline
    $Block_{header}$ & Bytes & \begin{tabular}{c} Data size\\of a Blockheader \end{tabular} & 80 Bytes \\
    \hline
    $Block_{content}$ & MB & \begin{tabular}{c} Data size\\of TXs in a block \end{tabular} & Variable \\
    \hline
    $BlockCount$ & Block & \begin{tabular}{c} Block count\\in a blockchain \end{tabular} & Variable \\
    \hline
  \end{tabular}
\end{table}

Parameters for estimating storage is shown in Table~\ref{table:StorageParameters}. In Bitcoin, the largest block size is 1MB. In Block data structure, Blockheader occupies 80 bytes.  The size of list of transactions included in a block is described as $Block_{content}$. Block count in the blockchain is $BlockCount$. The size of blockchain $BlockchainSize$ is shown as follows;
\begin{eqnarray}
  Block = Block_{header} + Block_{content}\\
  BlockchainSize = BlockCount \cdot BlockSize
\end{eqnarray}
As described in~\ref{VerifyTXandStorage}, an SPV node hold block headers. Therefore, the storage capacity for an SPV node $BlcokchainSize_{SPV}$ is like follows;
\begin{eqnarray}
  BlockchainSize_{SPV} = BlockCount \cdot Block_{header}
\end{eqnarray}

KARAKASA node distributedly store the whole blockchain. As a result, when operators want to deploy $N$ nodes, the required storage capacity is only $Blockchain$. In the case of Full Nodes, the capacity is as follows;
\begin{eqnarray}
  BlockchainSize \cdot N
\end{eqnarray}
In the case of SPV nodes, the capacity is as follows;
\begin{eqnarray}
  BlockchainSize_{SPV} \cdot N
\end{eqnarray}
Therefore, KARAKASA scheme required storage capacity lower than Full Nodes. Additionally, storage based on DHT is scalable. Thus, the KARAKASA scheme has an advantage that the required storage capacity does not become larger even with nodes scaling. Theoretically, the total storage capacity for KARAKASA is lower than SPV nodes when $N$ is over $1MB / 80Bytes = 13421772$. However, DHTs have not been verified to work on those number of nodes.

\subsection{Independent Verification of Transactions and Blocks}
In this section, we discuss the independent verification of transactions and Blocks on each scheme. Full Nodes are independent because Full Nodes can process verification by referring to the whole blockchain in the local storage. Therefore, Full Nodes do not need to trust other nodes. On the other hand, an SPV node depends on Full Nodes. Thus, an SPV node cannot independently verify transactions and blocks.

With the proposed scheme, a KARAKASA node can verify a new transaction using UTXOset in the local possession. In historical transaction and block verification, the node query the required information to KARAKASA cluster. The query happens when a KARAKASA node tries to resolve a fork. KARAKASA scheme adopts DHT to KVS of BlockStorage in a Bitcoin node. When a KARAKASA node query to the DHT cluster is the same as a Full Node queries to the local storage. Therefore, verification processes of transaction and block are also same. In other words, a KARAKASA node does not need to trust other nodes outside of the KARAKASA cluster.

As described in Section~\ref{TXrewriting}, if a node in a KARAKASA cluster rewrites a transaction, other nodes can detect whether the transaction is rewritten or not. Additionally, other nodes can get the transaction that has not been modified from other nodes that keep replicas. Therefore, by adopting schemes securing DHT cluster, there are no need to establish trusts in the DHT cluster. The trust problem of KARAKASA scheme is the same as the trust of storage of a Full Node. It means that the Full Node needs to trust the local storage. If an attacker rewrites the local storage of a Full Node, the Full Node can detect the rewriting as same as described in Section~\ref{TXrewriting}.

To secure DHT, it needs replications. On the other hand, it can also be considered that Full Nodes are replicating the whole blockchain for security. Thus, the replication in a KARAKASA cluster is the same as Full Node. Replication in KARAKASA scheme is achieved by the storage layer. On the other hand, Full Nodes replicates the whole node. In Section~\ref{StorageSizeForANode}, we described that KARAKASA scheme has a trade-off between replication and storage capacity. Therefore, there is a trade-off between security and storage capacity. By adopting the proposed scheme, operators can balance security and storage capacity on storage resource-constrained devices.

\section{Summary of This Chapter}
In this chapter, we compared KARAKASA node, Full Node, and SPV node. We compared from two points of view described in Chapter~\ref{viewpointofevaluation}. As the result of comparison of storage capacity, a KARAKASA node requires less storage capacity than a Full Node. Compare with an SPV node, when a KARAKASA cluster consists of over 13421772 nodes, required storage capacity for a KARAKASA node is smaller than an SPV node. From the viewpoint of the independent verification, KARAKASA node can independently verify transactions and blocks. The independent verification is achieved by utilizing characteristics and security schemes of DHT. Then, we conclude that KARAKASA node achieved the requirement described in Chapter~\ref{issue}. In the next chapter, we introduce some related works of KARAKASA.
\chapter{Related Works} 
\label{related} 
In this chapter, we introduce some related works of KARAKASA. There are some works evaluate blockchain systems~\cite{Gencer2018, Gervais2016, Bartoletti2017}. Applying those scheme, we plan to evaluate the effect of the difference of DHT algorithms. We also need to find suitable DHT algorithm for KARAKASA scheme. 

\section{Adopting DHT to Bitcoin and Blockchain}
There are several proposals applying DHT for blockchain. Kaneko et al. proposed a DHT clustering scheme for load balancing considering blockchain data size~\cite{kanekoDHT}. In their scheme, Bitcoin nodes are classified into two types; Mining nodes and data nodes. Data nodes verify transactions and blocks while holding the whole blockchain among a DHT cluster. Mining nodes work on mining with collecting transactions that are verified by data nodes. Data nodes are connected with each other by a DHT structured network. Data nodes are connected also by an unstructured network of Bitcoin. Since the verification of transactions is processed only by DHT networked data nodes, the propagation of transaction is processed rapidly according to the DHT network configuration. Comparing with KARAKASA scheme, in their scheme, the verification works are processed on a part of Bitcoin nodes. All KARAKASA nodes can verify all transactions and blocks by retrieving required data from the KARAKASA cluster. The independent verification is an advantage of KARAKASA scheme.

Frey et al. proposed that reducing storage capacity of an SVP node by mapping UTXOset in a DHT cluster~\cite{Frey2016}. Their proposal makes SVP nodes to verify transactions locally. They sharded UTXOset on DHT. Additionally, they embedded a hash of the hash list of blocks and a hash of shard onto a block data. With their scheme, each node is able to verify whether others rewrite the assgined part of UTXOset or not. KARAKASA scheme shards BlockStorage, not UTXOset. Their proposal focused on the reduction of storage capacity of SPV nodes. Thus, same as existing SPV node, their proposed SPV nodes need to rely on Full Nodes. In KARAKASA scheme, KARAKASA nodes does not need to rely on other nodes. Therefore, KARAKASA nodes keeps the independent verification even though keeping a whole blockchain distributedly.

Benshoof et al. proposed Distributed Decentralized Domain Name Service(DDDNS)~\cite{dddns}. Their work is based on~\cite{ddns1,ddns2,ddns3} that implements DNS on DHT. They applied a blockchain to DNS to authentication of record of domain name ownership. In their work, they suggested using the DHT overlay as the blockchain network overlay.

\section{Light Weight Nodes and Sharding}
Dorri et al. applied Blockchain technology to Internet-of-Things (IoT)~\cite{ali1,ali2,ali3}. They proposed Lightweight Scalable Blockchain (LSB)~\cite{LSB}. LSB is optimized for smart home architecture. In the smart home, resources of smart devices are limited on bandwidth and latency. The authors made a centralized manager to manage all incoming and outgoing request from resource-constrained devices. On the other hand, this thesis focuses on the reduction of storage capacity. It is possible to apply our scheme to smart devices to make local blockchain node in the smart home. Then, we can guarantee the integrity of data from smart devices and a centralized manager in the home while achieving better availability.

OmniLedger is one of the implementations of blockchains~\cite{OmniLedger,OmniLedger2,Sharding}. OmniLedger is scalable on the process on Full Node by sharding. The verification process of the transaction is sharded among some nodes. Ethereum is also planned to apply sharding~\cite{ethereumyellow, ETHsharding}. In KARAKASA scheme, the verification process is executable by every node in the cluster. To make Bitcoin scalable, sharding verification process is also essential. As one of our future work, we would optimize the process as well as storage capacity.

\section{DHT Applications and Evaluations}
Katari et al. evaluated and compared some replication schemes~\cite{Katariper2007}. They classified replication schemes into three patterns. Firstly, ``Neighbor Replication'' that replicate a data to a neighbor of the node assigned. Secondly, ``Path Replication'' replicate a data to a node on routing path look up. Thirdly, ``Multi Publication Key Replication'' that the data is assigned some IDs and is stored in some node that each ID assigned. Each scheme has different features. Landsiedel et al. evaluated fault tolerance of DHTs that replicated some scheme in churn condition~\cite{Landsiedel2006}. They described that more replication in the cluster, more strong fault tolerance. We evaluated KARAKASA scheme by simulation by referring current parameter of the Bitcoin network. We need to evaluate churn tolerance of KARAKASA scheme along the Bitcoin scenario.

DHTs provide efficient lookup of key-values in distributed nodes. There are several works evaluate DHTs~\cite{JinyanoLi, Varyani2016}. Storage based on DHT is structured on the overlay network. Hence, when nodes lookup a value, nodes communicate with each other to find the value. In this situation, latency and message loss may be significant. It is affected by real network topology, how a device connecting to the network and so on. We need to apply those works to Bitcoin scenarios. For example, we should consider how latency is forgiven for resource-constrained devices Bitcoin client.

\section{Summary of This Chapter}
In this chapter, we introduced related works of KARAKASA. The proposed scheme adoption DHT to Bitcoin and blockchain is that lightweight nodes rely on Full Node while load-balancing storage capacity. The advantage of KARAKASA is that nodes can behave independently. Other lightweight node proposals focus on sharding verification process of transactions and blocks as well as storage capacity. We can consider a combination of KARAKASA and those schemes for scale-up of Bitcoin performance. In adopting DHT to applications, there are several studies about replication and performance evaluation of DHT. We should consider the more practical evaluation of KARAKASA by adopting those studies. In the next chapter, we conclude this thesis and point out some future works of KARAKASA.
\chapter{Conclusion} 
\label{conclusion}
In this chapter, we conclude the thesis. We describe summery of the proposed scheme KARAKASA, and contoribution of the scheme. Finally, we point out some future works for KARAKASA scheme.

\section{Summery of the Proposed Scheme} 
This thesis proposed KARAKASA scheme that is a storage load balancing scheme among DHT clustered nodes. To independently verify transactions and blocks, existing schemes need to hold the whole blockchain. Otherwise, a node needs to rely on a Full Node. KARAKASA scheme made a node available to verify transactions and blocks independently with reducing the required storage capacity. The simulation of the KARAKASA scheme showed the size assigned a KARAKASA node was reduced. Security analysis showed that a node could detect rewriting assigned transactions in a malicious node in the KARAKASA cluster. Additionally, replication can guarantee the availability of a block. Therefore, KARAKASA nodes also keep independent verification of transaction and blocks. 

\section{Contribution} 
By KARAKASA scheme, an operator of a storage resource-constrained node can verify a Bitcoin transaction is processed correctly or not. Nowadays, the cryptocurrency exchange service provides a Full Node as a trustable endpoint. However, the benefit of Bitcoin is that a node that a user uses as an endpoint can independently verify transactions and blocks. The scheme of cryptocurrency exchange services requires users to trust their service. Therefore, the scheme of cryptocurrency exchange services is contradicting to Bitcoin idea. Ideally, all nodes in the Bitcoin network should be Full Nodes. However, the blockchain scheme has append-only characteristics. Therefore, estimating enough storage capacity is impossible.  The KARAKASA nodes can verify transactions and blocks independently. Hence, users can operate a KARAKASA node as an endpoint even though a storage resource-constrained devices.

\section{Future Works} 
At the end of this thesis, we point out some future works of KARAKASA scheme. In this decade, Bitcoin scaling problem has been on discussions. With resolving future works, KARAKASA scheme would be one of the solutions.

\subsection{Messaging Overheads} 
In Section~\ref{MessagingOverhead}, we pointed out the messaging overhead of building UTXOset would be linear to BlockCount. When the BlockCount becomes too large, the overhead would be large. From append-only characteristics of the blockchain, it is inevitable. Hence, KARAKASA scheme, just only distributed store based on DHT, is not complete. As a future work, we should consider the messaging overhead. The key element of the future work is the routing table and efficiency of DHT. We should discuss which DHT architecture is suitable for KARAKASA scheme with Bitcoin scenarios.

\subsection{Relationship between Security and Replication} 
In Section~\ref{Security}, we pointed out enough replication would be essential to make KARAKASA secure. How replication is enough for DHT is discussed~\cite{ghodsi2007symmetric, valduriez2004data}. We consider that we can determine the security parameter from how fault tolerance is required. The security of Bitcoin node is on the discussions~\cite{8369416}. We should consult those works to discuss the security of KARAKASA.

\subsection{Bitcoin Ecosystem with KARAKASA} 
In association with a security discussion of cryptocurrency exchange services, we should discuss where should set trust endpoint on Bitcoin. The endpoint needs to hold the whole blockchain. Adopting KARAKASA scheme, storage resource-constrained devices can be endpoints. The level of trust on endpoints is important. For example, we should discuss where to place the endpoint; at each user's devices, or each home, at each area. Nowadays, users should choose a Full Node, an SPV node, a client of cryptocurrency exchange services. The latter two depend on a trusted third party. A KARAKASA node is one of the new other options. The storage capacity required for the KARAKASA node was reduced. However, it is still larger than an SPV node. In another aspect, KARAKASA node is independently same as Full Nodes. There is no trust anchor to other nodes with enough replications. Comparing with SPV nodes, the independent verification is an advantage. A Bitcoin ecosystem design with those node types is one of future works. 

\clearpage

\bibliographystyle{unsrt}
\bibliography{chike_mthesis}
\addcontentsline{toc}{chapter}{References}
\renewcommand{\thechapter}{\Alph{chapter}}
\setcounter{chapter}{0}

\clearpage
\fancyhead[RO,LE]{ACKNOWLEDGEMENTS}
\chapter*{Acknowledgements}

First of all, I would thank Professor Dr. Jun Murai for a great contribution to today's Internet infrastructure,  teaching me "Research what you have LOVE.", and giving me the chance that I have interesting to Bitcoin and Blockchain. I would also be grateful for Professor Dr. Osamu Nakamura for his support to my research life. I also thank Project Associate Professor Dr. Shigeya Suzuki for discussions deeply about not only blockchain but also research and my life. I could not complete writing this thesis without his grateful support. Moreover, special thanks for faculties of Mobile and Ubiquitous Internet Project, Professor Dr. Kazuto Ataka, Associate Professor Dr. Keisuke Uehara, Professor Dr. Hiroyuki Kusumoto, Professor Dr. Keiji Takeda, Associate Professor Dr. Rodney D. Van Meter, Professor Dr. Jin Mitsugi, Project Associate Professor Dr. Masaaki Sato, Project Assistant Professor Dr. Achmad Husni Thamrin, Project Assistant Professor Dr. Yohei Kuga, Project Assistant Professor Dr. Takeshi Matsuya.

I have had the support and encouragement of Jun Murai Lab. BCALI group a.k.a ``Blockchain Boys.''; Mr. Sotaro Ichiyama, Mr. Asato Ueda, Mr. Yuta Morino, Mr. Yuya Murakami, Mr. Koki Hirokawa, Mr. Fumihiro Kinoshita, Mr. Hideki Komaki, and Mr. Koki Kawano. Their toughness and interest to Blockchain strongly boosted me to research.

I would also thank my colleague of Jun Murai Lab. Mr. Kohei Suzuki, Mr. Florindo da Costa a.k.a ``Blockchain Guys'' discussed blockchain technology with saying ``Watashi no Jinsei Owari.'' Ms. Mariko Kobayashi, Ms. Yuka Sasaki, Ms. Haruka Nakajima (a.k.a ``IoT girls''), and Ms. Rina Ueno encourage me from other research fields. They fought to write each master thesis with me. 

I thank seniors of Jun Murai Lab. Dr. Shota Nagayama gave me advice for research activity and my life. Mr. Hirotaka Nakajima gave me chance to join internship in Merpay, Inc. Mr. Yusaku Sawai, Mr. Shota Kawamoto, Mrs. Keiko Shigeta Kato, Mr. Kohei Yamamoto, Mr. Iori Mizutani tell me the severity of master course. 

Special thanks also to juniors of Jun Murai Lab. Mr. Yuta Sugafuji smoked too many with me. Mr. Yasunobu Toyoda supported to deploy the experiment environment. Moreover, many juniors, Mr. Shuya Ozaki, Mr. Tao Oshimi, Mr. Daiki Kamei, Mr. Ryoma Kawaguchi, Mr. Sena Kuwabara, Mr. Akira Shoji, Mr. Hiromu Kamei, Ms. Asako Seshimo, Mr. Kei Nomaguchi, Mr. Kazuaki Masuda, Mr. Yuichi Masda, Ms. Ayana Yoroisaka, Mr. Richard Rowland, entertained me the Laboratory life.

I also want to thank gratefully to NTT Service Evolution Laboratory Blockchain team members; Mr. Hiroki Watanabe, Mr. Shigenori Ohash, Mr. Shigeru Fujimura, Dr. Atsushi Nakadaira. My internship on the team was a very important experience for my life. I would want to discuss blockchain and future from now on.

I also thank to members of Mercari, Inc. and Merpay, Inc. Blockchain team; Mr. Misato ``Misatoken'' Takahashi, Mr. Keita Nakamura, Mr. Kentaro Teramoto, Ms. Paya Do, Mr. Agro Rachmatullah, Mr. Yusuke Shimizu, Mr. Kenji Kubo, Mr. Yasuo Kobayashi, Mr. Seiyo Kurita, and Mr. Tomoya Ishizaki. Their view of blockchain and its futures realized me the importance of bridging between academic researches and industrial products. 

My deepest heartfelt appreciation goes to The Mitsubishi UFJ Trust Scholarship Foundation, and KEIO ENGINEERING FOUNDATION for the scholarship. For their support, I could concentrate on research activity. Additionally, I also special thanks for Keio University SFC Student Life Department for supporting my submission of scholarships. Especially, I'm grad to Mr. Norihiro Ida and Mrs. Keiko Ohta for politely and carefully responded.

I would also thank my friends. Mr. Toshiki Hosokawa drank with me and discussed life, research, work, and so many topics. Mr. Yuta Hasegawa and Mrs. Maria Hasegawa gave me refresh time with my wife and supported my marriage life. Mr. Fumiya Kato and Ms. Ai Kutsuzawa gave me advice for my life. 

I also thank the musicians I love; Maximum The Hormone, Heresy, Layout my Torturechamber, Slipknot, a crowd of rebellion, THE BACK HORN, and so on. Their music encouraged me through my research life.

I would say thanks to Philip Morris International Inc. Their greatest products ``IQOS'' and ``Marlboro'' supported my relaxing smoking time while writing this thesis.  

I'm grateful to the environment, all Faculty members and everything of Keio University SFC. Nature rich environment and people that interested in various topics gave me many stimuli. 

I would like to show my greatest appreciation to my family; grandfather Hiroyuki Kokubo, grandmother Motoko Kokubo, mother Hiroko, father Yuichi, brother Shunsuke, and sister Fukiko. They supported my life greatly.

Finally, I owe my deepest gratitude to my beloved wife Ayaka, and my daughter Ibuki. Their grateful support gave me power. They made a relaxing time for me with our pet dog, LuLu, and CoCo. My life got a better one towards futures.

\addcontentsline{toc}{chapter}{Acknowledgements}
\clearpage


\end{document}